\documentclass[sigconf,authorversion,table]{acmart}

\usepackage[T1]{fontenc}
\usepackage[utf8]{inputenc}
\usepackage{hhline}

\AtBeginDocument{%
  \providecommand\BibTeX{{%
    \normalfont B\kern-0.5em{\scshape i\kern-0.25em b}\kern-0.8em\TeX}}}

\copyrightyear{2020}
\acmYear{2020}
\setcopyright{acmlicensed}
\acmConference[I3D '20]{Symposium on Interactive 3D Graphics and Games}{May 5--7, 2020}{San Francisco, CA, USA}
\acmBooktitle{Symposium on Interactive 3D Graphics and Games (I3D '20), May 5--7, 2020, San Francisco, CA, USA}
\acmPrice{15.00}
\acmDOI{10.1145/3384382.3384534}
\acmISBN{978-1-4503-7589-4/20/05}

\newcommand{\compress}{\vspace*{-2mm}}
\newcommand{\plotwidth}{0.42}

\newcommand{\MCPS}{$\overline{M}^{64}_{CPS}$~}

\newcommand{\rev}[1]{#1}

\begin{document}

\title{On Ray Reordering Techniques for Faster GPU Ray Tracing}

\author{Daniel Meister}
\affiliation{%
  \institution{The University of Tokyo}
  \institution{Czech Technical University in Prague}}
\additionalaffiliation{Japan Society for the Promotion of Science as International Research Fellow}

\author{Jakub Bok\v{s}ansk\'{y}}
\affiliation{%
  \institution{NVIDIA}
  \institution{Czech Technical University in Prague}}

\author{Michael Guthe}
\affiliation{%
  \institution{The University of Bayreuth}}

\author{Ji\v{r}\'{\i} Bittner}
\affiliation{%
  \institution{Czech Technical University in Prague}}

\renewcommand{\shortauthors}{Meister, Bok\v{s}ansk\'{y}, Guthe, and Bittner}

\begin{abstract}
We study ray reordering as a tool for increasing the performance of existing GPU ray tracing implementations. We focus on ray reordering that is fully agnostic to the particular trace kernel. We summarize the existing methods for computing the ray sorting keys and discuss their properties. We propose a novel modification of a previously proposed method using the termination point estimation that is well-suited to tracing secondary rays. We evaluate the ray reordering techniques in the context of the wavefront path tracing using the RTX trace kernels. We show that ray reordering yields significantly higher trace speed on recent GPUs ($1.3-2.0\times$), but to recover the reordering overhead in the hardware-accelerated trace phase is problematic.
\end{abstract}

\ccsdesc[500]{Computing methodologies~Ray tracing}
\ccsdesc[500]{Computing methodologies~Visibility}
\ccsdesc[500]{Theory of computation~Sorting and searching}

\keywords{ray tracing, ray sorting, real-time rendering, RTX}

\begin{teaserfigure}
\setlength{\tabcolsep}{1pt}
\begin{tabular}{ccc}
 \includegraphics[width=0.33\linewidth,trim=200 0 200 0,clip]{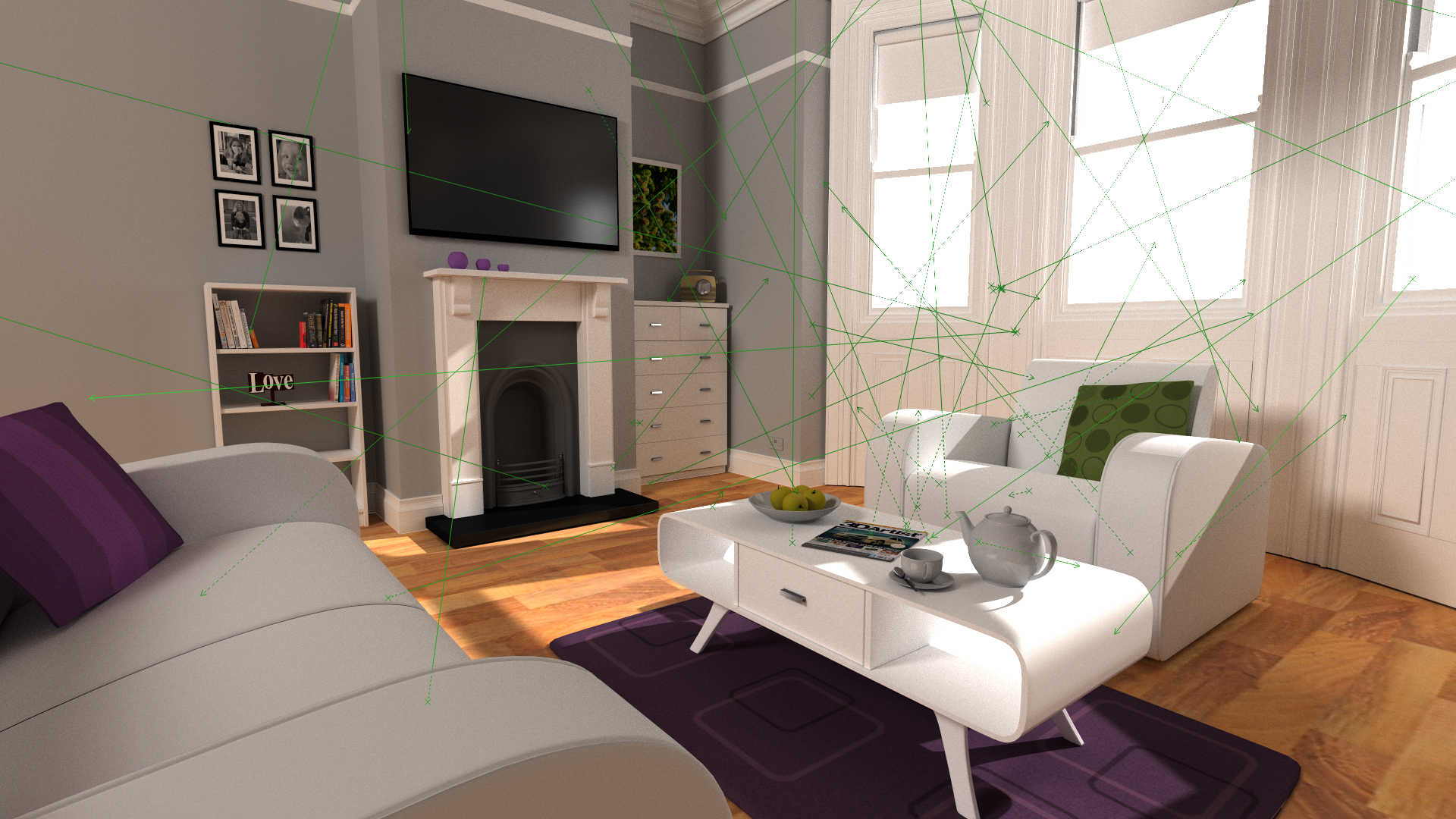} &
 \includegraphics[width=0.33\linewidth,trim=200 0 200 0,clip]{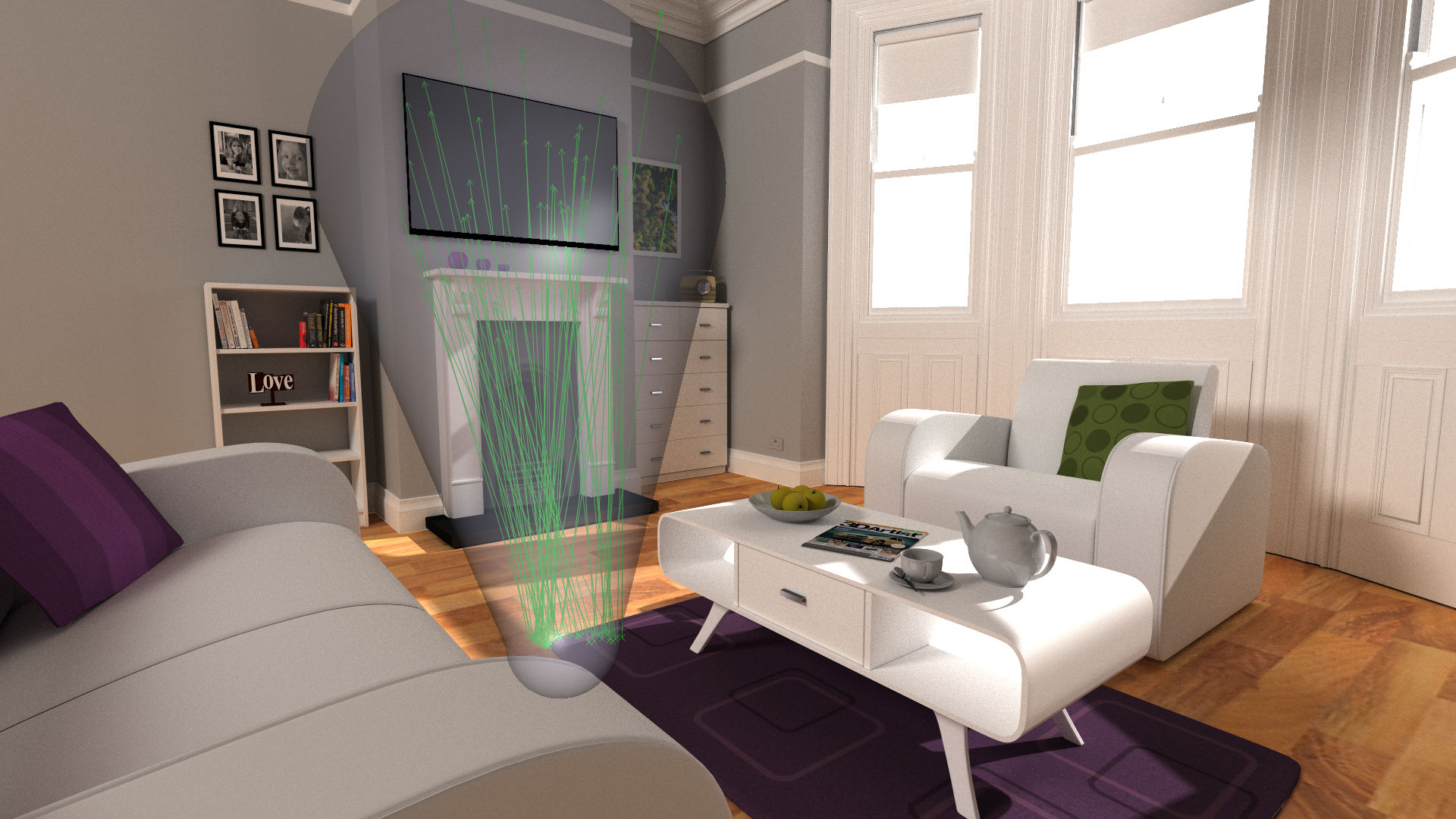} & 
 \includegraphics[width=0.33\linewidth,trim=200 0 200 0,clip]{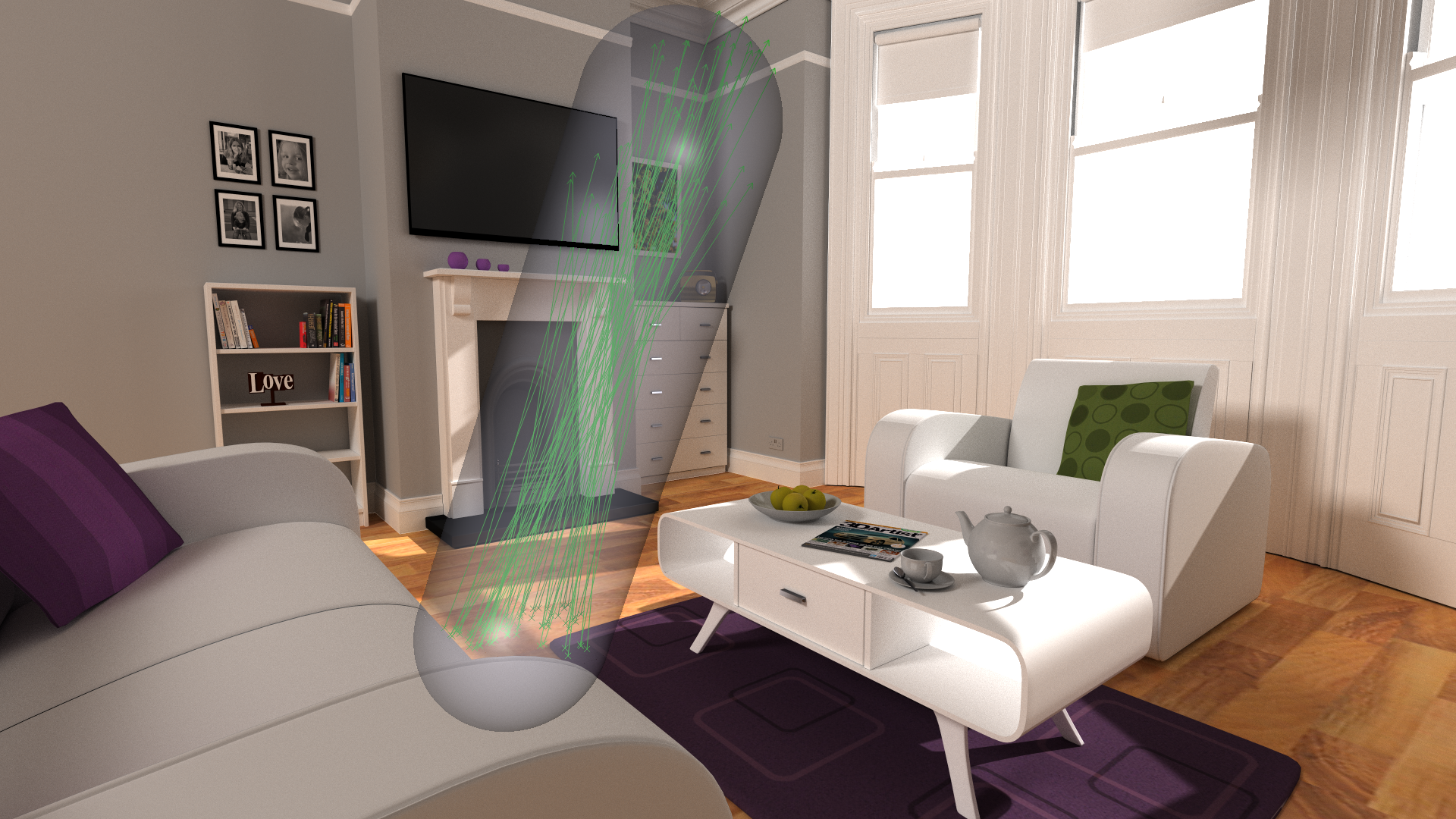}\\ 
2355 MRays/s & 3593 MRays/s & 3914 MRays/s
\end{tabular}
  \caption{Visualization of a subset of 64 consecutive rays from the third bounce of the wavefront path tracer. The numbers below the images show trace performance for that bounce on the RTX 2080 Ti GPU using RTX trace kernels: original incoherent rays (left), rays reordered using origin-direction sorting key (middle), rays reordered using the origin-termination sorting key (right)}
  \label{fig:teaser}
\end{teaserfigure}

\maketitle

\section{Introduction}
Ray tracing is a popular rendering paradigm that allows us to simulate light propagation by tracing rays and reconstructing the corresponding light paths. Many rays have to be traced to compute a high-quality image. The initial ray tracing methods, such as Whitted style ray tracing, use rays that exhibit a high degree of coherence within a given stage of the algorithm. However, the currently employed techniques such as path tracing produce increasingly incoherent ray sets due to scattering on diffuse and glossy surfaces. Tracing incoherent rays is much more costly than tracing coherent ones due to higher memory bandwidth, higher cache miss rate, and computational divergence. A number of techniques were proposed to mitigate this issue that usually use the wavefront path tracing combined with ray reordering, packet tracing, or ray space hierarchies. 

The core of the ray tracing based algorithms is evaluating ray scene intersections, which is often referred to as the trace kernel. In our paper, we revisit the basic problem of sorting rays to produce coherent subsets of rays in order to accelerate the trace kernel. We focus on methods that are fully agnostic to the particular trace kernel and the employed acceleration data structure. Such techniques already appeared in the literature~\cite{Aila2010,Costa2015,Moon2010,Reis2017}, but we feel there is a need for their thorough comparison and deeper analysis.

\noindent
We aim at the following contributions:
\begin{itemize}
\item We summarize previously published methods for ray reordering suitable for GPU ray tracing.
\item We propose a method for sorting key computation that aims to maximize ray coherence by using a novel termination point estimation technique.
\item We show the current limits of the trace acceleration using an idealized ray reordering algorithm for the RTX trace kernels.
\end{itemize}

\section{Related Work}
Global illumination is one of the so-called embarrassingly parallel problems since it performs the same algorithm for each sample of every pixel. Finding the closest intersection during the trace phase for incoherent rays, however, leads to thread and data access divergence, which drastically reduces throughput~\cite{Wald2003}. One possibility to handle this issue is explicitly generating coherent ray sets. For this, \citet{Szirmay1998} proposed global ray bundle tracing; and more recently, \citet{Nimier2019} used coherent MCMC sampling. These methods can generate highly coherent workloads with minimal overhead, but they require a complete redesign of the underlying rendering algorithm.

\citet{Pharr1997} proposed to reorder the traversal such that all rays are tested only against a subset of the scene to improve cache coherence. While this is viable for very large scenes, repeatedly loading ray data significantly increases the memory bandwidth. \citet{Navratil2007} extended this idea by interpreting intersection testing as a scheduling problem and proposed to split both rays and geometry into cache blocks.

For production rendering, not only the trace kernel but also shading might be limited by memory bandwidth. Therefore, \citet{Eisenacher2013} proposed to sort termination points to improve shading performance. While this approach is designed for out-of-core path tracing, grouping shading calculations by a material also improves in-core performance for complex shaders. \rev{For highly detailed scenes, \citet{Hanika2010} proposed to use a two-level hierarchy combined with ray sorting to facilitate efficient on the fly micro-polygon tessellation. The rays are traversed through the top-level hierarchy, and they are repeatedly sorted to determine sets of rays traversing the same leaf nodes of the top-level hierarchy.}

When coherence among rays exists, the packet traversal~\cite{Gunther2007} exploits it by forcing a SIMD processing of a group of rays. This, on the other hand, increases inter-thread communication and synchronization. Furthermore, it assumes high ray coherence and is significantly slower than depth-first traversal for incoherent rays. \citet{Bikker2012} proposed a packet traversal algorithm that uses batching to improve data locality.

Current state-of-the-art GPU-based global illumination frameworks are based on splitting the work into calculating the intersection of rays with the scene (the trace kernel) and shading~\cite{Laine2013}. The fact that the ordering in which rays are processed in the trace kernel can be independent of shading makes ray sorting or reordering approaches feasible.

In a case when thread divergence occurs on GPU, the whole warp of threads is blocked until all its rays finish the traversal. \citet{Aila2009} proposed to increase SIMD efficiency by replacing already finished rays with new ones from a global queue. Techniques such as speculative traversal slightly increase the redundancy of ray intersection tests because they work on possibly terminated rays. But this redundancy is only virtual as the core would otherwise be idle. \citet{Boulos2008} proposed a packet reordering for SIMD ray tracing on CPU. Analogously, thread compaction algorithms~\cite{vanAntwerpen2011,Wald2011}, which remove empty secondary rays, were proposed for GPU ray tracing. Although the number of processed warps is reduced by the factor of up to 5, the performance only increases by up to 16\% due to increased divergence.

\citet{Garanzha2010} used breadth-first packet traversal after a ray sorting step. They proposed the idea of sorting rays to reduce divergence in computation using a hash-based method for sorting the rays into coherent packets. This method was inspired by the approach of \citet{Arvo1987} for CPU ray tracing. In addition, the rays are grouped into frusta, which are further tested against the scene as proposed by \citet{Reshetov2005}. This way, the total number of intersection tests is reduced. While they report impressive speedups for primary rays and deterministic ray tracing, this does not translate to path tracing because the frusta become too large and intersect most of the scene.

When analyzing the efficiency of ray tracing on GPUs, \citet{Aila2009} also evaluated a hash-based sorting criterion based on interleaving ray origin and normalized ray direction. At that time, the sorting overhead was too large to improve the overall rendering time. Another ray reordering scheme addressing cache coherent memory access for out-of-core rendering is used by \citet{Moon2010}. They propose to sort rays using an estimated termination point that is calculated by ray tracing a simplified scene that fits into the main memory. The approach is, however, only suitable for out-of-core ray tracing due to the expensive hit point estimation. More recently, \citet{Costa2015} proposed a ray sorting approach for ambient occlusion. Their sorting criterion is, however, mainly based on ray direction and is suitable for short rays or shadow rays, where the direction depends on the origin.

Considering the problem of divergent rays in most of the global illumination algorithms, \citet{Aila2010} proposed a hardware implementation specifically designed to trace incoherent rays. Comparing to the RTX design, however, it shows that today's hardware implementations are still optimized for coherent ray batches.

\section{Ray Reordering}
Rays in three-dimensional space can be represented as points in a five-dimensional space (ray space), where three dimensions represent ray origins, and two dimensions represent ray directions. The ray space can be embedded into six dimensions, either for the sake of uniformity of ray direction representation or to account for the ray length (e.g. for shadow rays). In that case, 3D rays form a subset of 6D space.

Reordering of incoherent rays to form a more coherent ray set can be achieved by different strategies. We can use a multidimensional divide-and-conquer strategy~\cite{Szecsi2006}, or various clustering techniques, such as the $k$-means clustering~\cite{Lloyd1982}. In our paper, we focus on techniques that use mapping of the multidimensional ray space to one-dimensional space of sorting keys, which are ordered using a standard efficient sorting algorithm. The sorting key computation discretizes the ray space into a set of multidimensional cells and uses a particular mapping of their coordinates to one-dimensional indices. The resulting indices form a space-filling curve, visiting all cells of the discretized ray space. In most cases, the Morton curve is used for simplicity of its calculation based on simple bit interleaving~\cite{Lauterbach2009,Aila2010,Costa2015}.

\newcommand{\dir}{\cellcolor{blue!25}d}
\newcommand{\orig}{\cellcolor{green!25}o}
\newcommand{\term}{\cellcolor{yellow!25}t}
\newcommand{\mlabel}[1]{\begin{small}#1\end{small}}

\begin{figure*}[t]
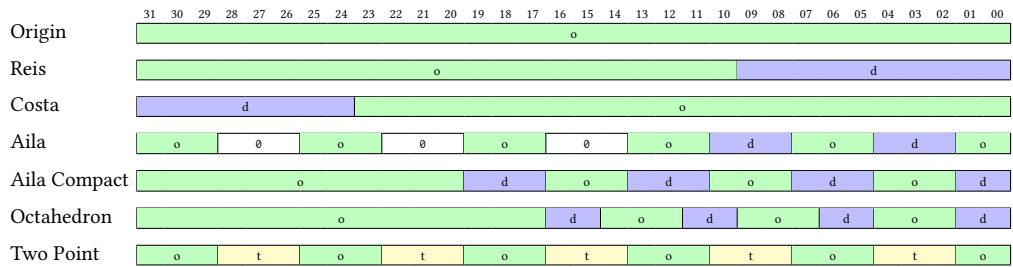

\begin{center}
\begin{tiny}
  \setlength\tabcolsep{1mm}
  \begin{tabular}{lcccccccccccccccccccccccccccccccc}
    & 31 &
    30 & 29 & 28 & 27 & 26 & 25 & 24 & 23 & 22 & 21 &
    20 & 19 & 18 & 17 & 16 & 15 & 14 & 13 & 12 & 11 &
    10 & 09 & 08 & 07 & 06 & 05 & 04 & 03 & 02 & 01 & 00 \\
    \hhline{~--------------------------------}
    \mlabel{Origin} & \multicolumn{32}{|c|}{\orig} \\
    \hhline{~--------------------------------}
    & \\
    \hhline{~--------------------------------}
     \mlabel{Reis} & \multicolumn{22}{|c|}{\orig} & \multicolumn{10}{c|}{\dir} \\
    \hhline{~--------------------------------}
    & \\
    \hhline{~--------------------------------}
     \mlabel{Costa} & \multicolumn{8}{|c|}{\dir} & \multicolumn{24}{c|}{\orig} \\
    \hhline{~--------------------------------}
    & \\
    \hhline{~--------------------------------}
     \mlabel{Aila} & 
    \multicolumn{3}{|c|}{\orig} & \multicolumn{3}{c|}{\texttt{0}} &
    \multicolumn{3}{c|}{\orig} & \multicolumn{3}{c|}{\texttt{0}} &
    \multicolumn{3}{c|}{\orig} & \multicolumn{3}{c|}{\texttt{0}} &
    \multicolumn{3}{c|}{\orig} & \multicolumn{3}{c|}{\dir} &
    \multicolumn{3}{c|}{\orig} & \multicolumn{3}{c|}{\dir} &
    \multicolumn{2}{c|}{\orig} \\
    \hhline{~--------------------------------}
    & \\
    \hhline{~--------------------------------}
     \mlabel{Aila Compact} & \multicolumn{12}{|c|}{\orig} &
    \multicolumn{3}{c|}{\dir} & \multicolumn{3}{c|}{\orig} &
    \multicolumn{3}{c|}{\dir} & \multicolumn{3}{c|}{\orig} &
    \multicolumn{3}{c|}{\dir} & \multicolumn{3}{c|}{\orig} &
    \multicolumn{2}{c|}{\dir} \\
    \hhline{~--------------------------------}
    & \\
    \hhline{~--------------------------------}
     \mlabel{Octahedron} & \multicolumn{15}{|c|}{\orig} &
    \multicolumn{2}{c|}{\dir} & \multicolumn{3}{c|}{\orig} &
    \multicolumn{2}{c|}{\dir} & \multicolumn{3}{c|}{\orig} &
    \multicolumn{2}{c|}{\dir} & \multicolumn{3}{c|}{\orig} &
    \multicolumn{2}{c|}{\dir} \\
    \hhline{~--------------------------------}
    & \\
    \hhline{~--------------------------------}
     \mlabel{Two Point} & 
    \multicolumn{3}{|c|}{\orig} & \multicolumn{3}{c|}{\term} &
    \multicolumn{3}{c|}{\orig} & \multicolumn{3}{c|}{\term} &
    \multicolumn{3}{c|}{\orig} & \multicolumn{3}{c|}{\term} &
    \multicolumn{3}{c|}{\orig} & \multicolumn{3}{c|}{\term} &
    \multicolumn{3}{c|}{\orig} & \multicolumn{3}{c|}{\term} &
    \multicolumn{2}{c|}{\orig} \\
    \hhline{~--------------------------------}
  \end{tabular}
\end{tiny}
\end{center}
\caption{Overview of different sorting key computation methods. The figure shows bits of a 32-bit key occupied by encoding origin (o), direction (d), estimated termination point (t), or zeros (0). The consecutive triples of o, d, and t bits use the x, y, z component ordering, which are omitted in the table for better readability.}
\compress
\label{fig:codes}
\end{figure*}

\subsection{Sorting Key Computation Methods}
The main aspect of the ray reordering algorithm based on space-filling curves is the particular method used to compute the sorting key. Several techniques for this step have already been proposed. We briefly summarize the previously published methods, and then we proceed to describe two novel techniques.

\subsubsection{Origin}
The simplest way of reordering the rays is to enforce the coherence of their origins. This can be achieved by using the origins as a sorting key to form a 3D space-filling curve. Resulting subsets have coherent origins, but their directions may vary greatly.

\subsubsection{Origin-Direction}
To improve ray coherence, we can include the direction into the sorting key using lower bits of the key for the parametrized direction. This method was proposed by \citet{Reis2017} for sorting secondary rays when constructing a ray space hierarchy. Sorting rays using this key leads to ray subsets that have coherent origins up to the given resolution specified by the number of bits allocated for the origin representation in the sorting key. If there are more rays with the same discretized origin, they will also be ordered by their direction (lower bits of the code). For rays with incoherent origins and directions, this technique improves ray coherence as the directional information is also reflected. On the other hand, the method is sensitive to the number of bits allocated for the origins. For example, the consecutive subsets of rays may cover the whole range of directions if too many bits were used for origins.

\subsubsection{Direction-Origin}
An alternative to the technique described above is to use the direction information in higher bits of the code followed by the bits representing the origin. This method was proposed by \citet{Costa2015} for sorting shadow and ambient occlusion rays. This technique shares the problem of allocating a reasonable number of bits for the component represented by more significant bits of the sorting key, in this case, the direction.

\subsubsection{Origin-Direction Interleaved}
Another strategy of computing sorting keys uses interleaving of bits representing origin and direction. This technique was used by \citet{Aila2010} to study the behavior of tracing incoherent rays. The advantage of this strategy is that the sorting key corresponds to a multidimensional space-filling curve that progressively encodes both origins and directions. Therefore it is not sensitive to the specification of the number of bits used for the origin and/or direction. The direction is represented by embedding a sphere into a 3D space using normalization of the directional component of the ray. The quantized direction is prefixed with leading zeros (three zero bits for each directional component). This effectively implies sorting based on origins on a coarser scale and then sorting by the direction and origins interleaved.

\citet{Aila2009} used the quicksort algorithm with 192-bit sorting keys on CPU in their publicly available implementations. For practical usage, the sorting key has to be shorter (64 bits at max.) that the sorting overhead can potentially be recovered during the trace phase. Shorter sorting keys reduce memory traffic, and also the number of internal passes of an underlying sorting algorithm such as bucket sort or radix sort.

An alternative to the encoding of Aila and Karras is to use a more compact representation of ray direction such as the octahedron parametrization~\cite{Meyer2010}. The octahedron parametrization achieves high uniformity and high locality of its mapping to the 1D space using the Morton curve. In our tests, we used the original sorting key proposed by Aila and Karras, as well as its compacted version that removes the leading zeros from the higher bits of the code. We also tested a novel sorting key that uses the octahedron direction parametrization. 

\subsubsection{Two Point Sorting Key}
To achieve compact bounding volumes of consecutive ray subsets, we should enforce their coherence by forming compact sets of ray origins as well as its termination points. Such subdivision leads to smaller overlap of disjoint ray subsets in space, and thus to more compact bounding volumes. The problem of this strategy is the requirement to know the ray termination point before computing the sorting key, i.e. the termination point is what we aim to compute by tracing the ray. 

The idea of exploiting the termination points was first coined by \citet{Moon2010} who proposed to use the estimated termination points as a primary sorting key for ray tracing out-of-core scenes. They used a termination point estimation using a simplified geometry of the scene. The authors also report results for combining the estimated termination point with origins or directions for sorting key evaluation without describing the actual details for the sorting key bit layout. We revisit this idea by proposing two different termination point estimations. We also describe the actual bit-layouts.

The Two Point method estimates the termination point and computes the sorting key using the 6D point which represents the ray. The key is constructed by interleaving bits of origin and termination points. Our layout slightly prioritizes ray origins as these are known precisely unlike the termination points. An overview of all aforementioned sorting key computation methods is shown in Figure~\ref{fig:codes}.

\subsubsection{Termination Point Estimation}
We propose two methods on how to estimate the termination points, i.e. to estimate ray lengths. The first method is setting a constant estimated ray length value across the scene using a fixed ratio of the largest extent of the scene bounding box. This method is very simple and provides surprisingly good results. We performed a large set of measurements, and the best overall setting was in the range of 0.2 to 0.3 of the largest scene extent. In the results section, we use the value of 0.25 for evaluating the fixed-length estimation variant of the Two Point method. \rev{Note that despite using a fixed-length estimation, the reordering results will not be the same as for the origin-direction method, in which the directional part of the sorting key solely depends on the ray direction independently of the ray origin}.

Using the constant value may be an issue for the teapot in the stadium scenes. Thus, we propose a second method based on adaptive caching of the ray length from the previous trace passes. We use a spatial hash table using short Morton codes as keys. In each cell of the hash table, we store the sum of ray lengths and the ray counts. After each trace pass, we accumulate new values into the hash table. To recover the ray length estimate, we simply compute a Morton code to get the corresponding cell and compute the average ray length using the values of the cell. Note that we have to use another method for computing Morton codes to access the hash table as we do not know the ray length at that moment. We use 20-bit Morton codes \rev{computed with the compacted version of Aila's method}, which correspond to the hash table with $2^{20}$ cells. To prevent getting unpredictable results by querying empty cells, we initialize each cell of the hash table with a dummy ray with length 0.25 \rev{of the largest scene extent}.

As a reference technique, we use the actual termination points evaluated by ray tracing for computing the sorting key. This technique is not a practical use case; however, it allows us to define the limits of the Two Point method if a precise termination point estimation was available.

\subsection{Sorting Algorithm}
\label{sec:sorting}
Sorting speed is crucial for any practical use of the ray reordering techniques. We use parallel radix sort proposed by \citet{Merrill2011} (a part of the CUB library), which is the fastest sorting algorithm to date, according to best of our knowledge. The authors of the algorithm report sorting speedup up to 6190 MKeys/s on Tesla P100. On the RTX 2080 Ti GPU and 32-bit sorting keys, it achieves 1600 MKeys/s for one million keys, 2100 MKeys/s for two million keys, and 2700 MKeys/s for 4 million keys. Therefore, sorting rays for resolution \rev{$1920\times1080$} with one sample per pixel takes about 1 ms. The sorting is performed on key-index pairs first, and then we use an additional pass to perform the actual ray data reordering.

We can sort the Morton codes either globally or locally (e.g. segments 1024 Morton codes) using a specialized kernel of the CUB library. The local sorting can be done more efficiently; however, it usually results in a significantly less coherent ray set.

\section{Results}
\begin{figure*}[htb]
  \centering
  \hfill
\includegraphics[width=0.49\linewidth]{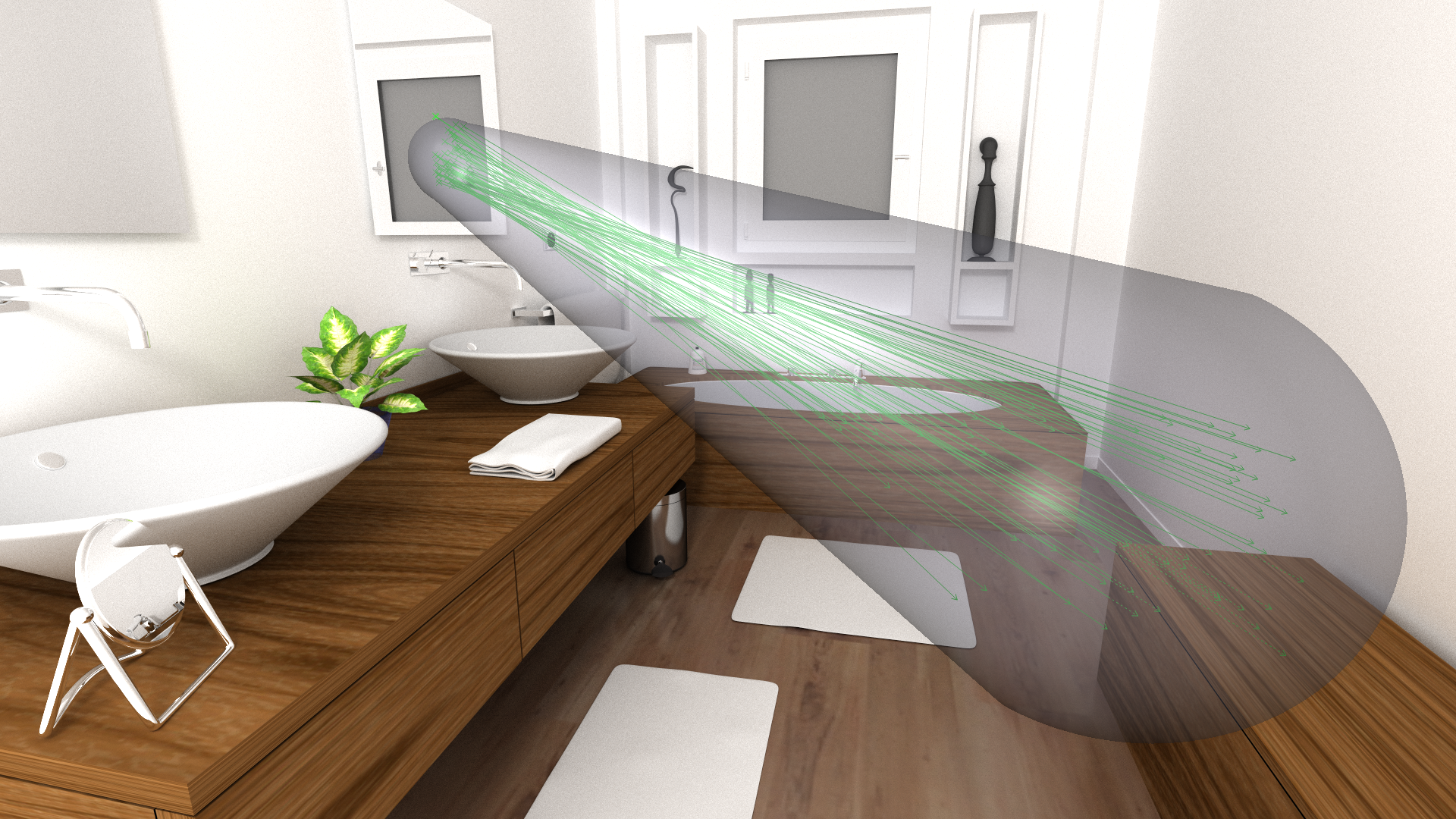}
  \hfill
\includegraphics[width=0.49\linewidth]{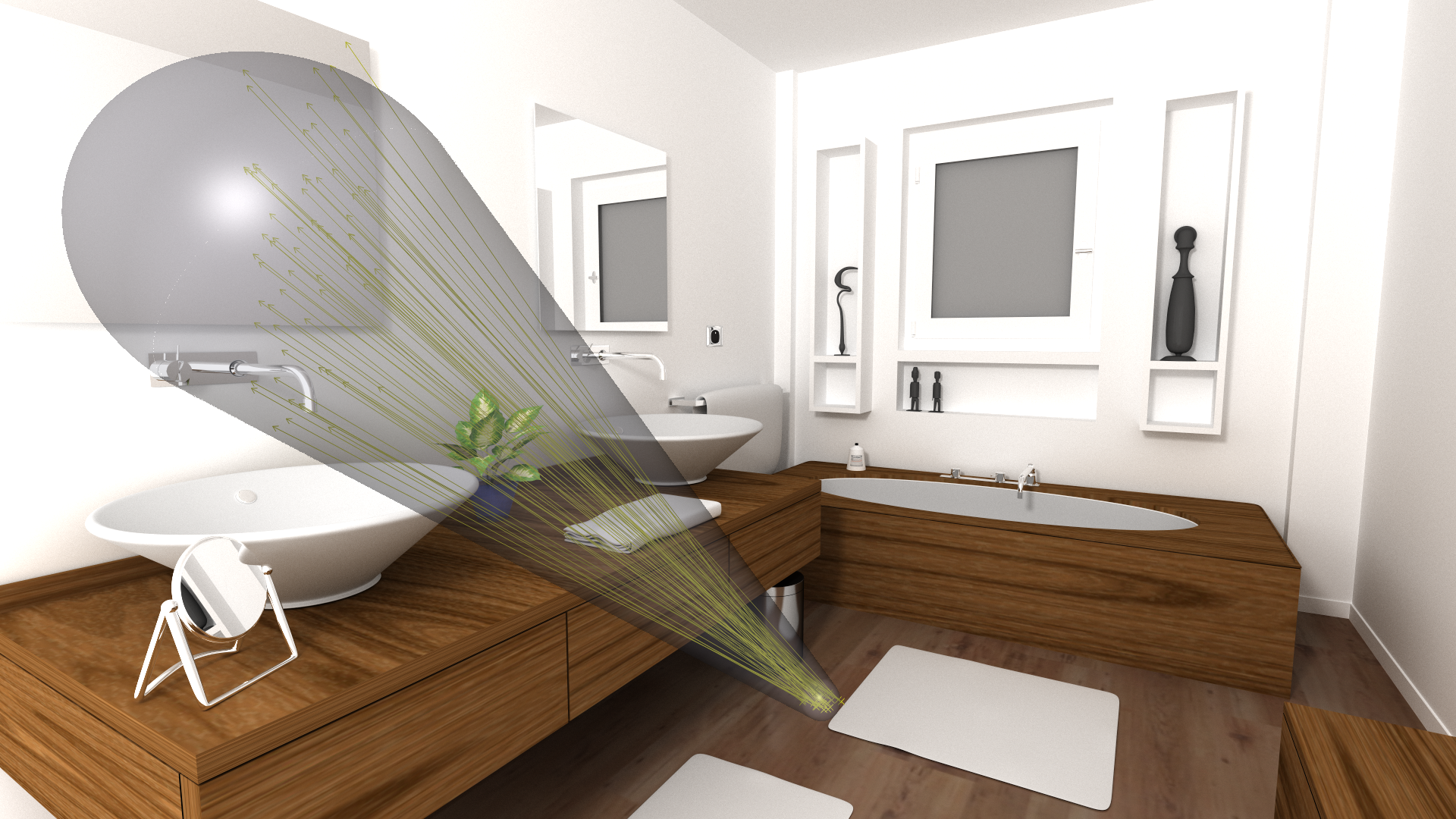}
  \hfill
   \caption{\label{fig:visualization} The coherent subsets of consecutive rays for the 6th bounce in the wavefront path tracer as a result of ray reordering using our Two Point method: secondary rays reflected from a mirror surface (left) and shadow rays cast towards a large area light (right).}
\end{figure*}

We evaluated the discussed ray reordering techniques in the context of wavefront path tracing using hardware accelerated RTX trace kernel accessed through DirectX 12 and OptiX 7. The path tracer uses next event estimation with two shadow rays per bounce with eight samples per pixel. We use seven scenes of various complexity \cite{Bitterli2016,McGuire2017} with a single area light source with the size of 5\% of the largest scene extent (see Figure~\ref{fig:visualization}). All measurements were performed on the RTX 2080 Ti GPU with the image resolution of $1920\times1080$. We measured the discussed ray ordering strategies in terms of trace performance, hardware utilization, and ray coherence measures.

\begin{table*}
\caption{Trace speed comparison of tested methods using 32-bit sorting keys for secondary and shadow rays (measured on the RTX 2080 Ti GPU).}
\begin{center}
\setlength{\tabcolsep}{3.5pt}
\scalebox{0.62}{
\begin{tabular}{lccccccccc}
\hline
\hline
& Crytek Sponza & Pub & Resort & Living room & Salle de Bain & Breakfast & Bistro &\\
& \raisebox{1em}[1.21\height]{\includegraphics[height=0.1\textheight,trim=400 0 400 0,clip]{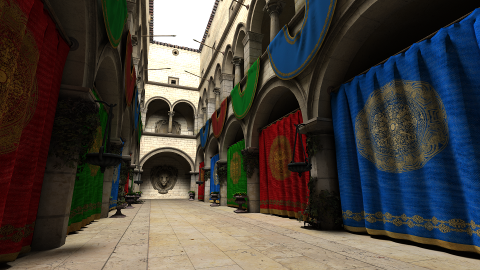}} & \raisebox{1em}[1.21\height]{\includegraphics[height=0.1\textheight,trim=400 0 400 0,clip]{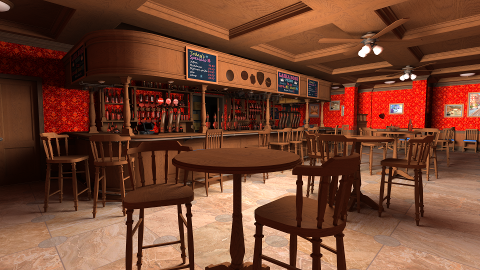}} &  \raisebox{1em}[1.21\height]{\includegraphics[height=0.1\textheight,trim=400 0 400 0,clip]{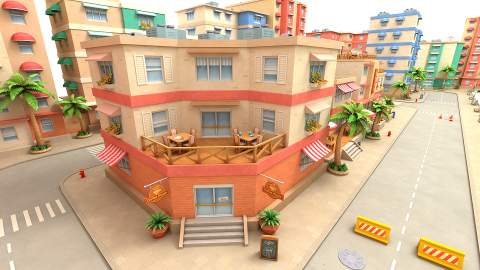}} & \raisebox{1em}[1.21\height]{\includegraphics[height=0.1\textheight,trim=400 0 400 0,clip]{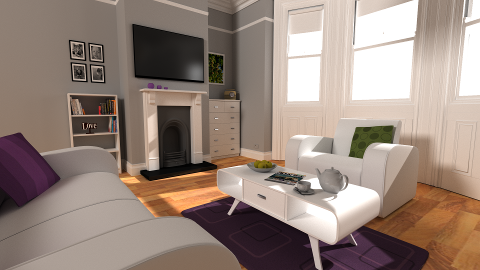}} & \raisebox{1em}[1.21\height]{\includegraphics[height=0.1\textheight,trim=400 0 400 0,clip]{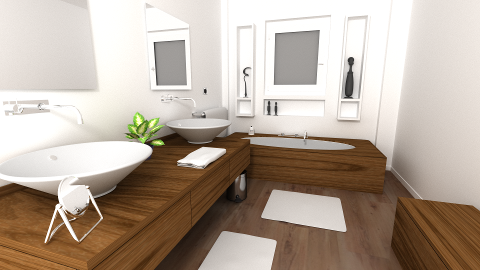}} & \raisebox{1em}[1.21\height]{\includegraphics[height=0.1\textheight,trim=400 0 400 0,clip]{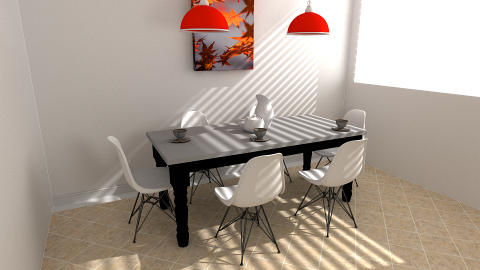}} & \raisebox{1em}[1.21\height]{\includegraphics[height=0.1\textheight,trim=400 0 400 0,clip]{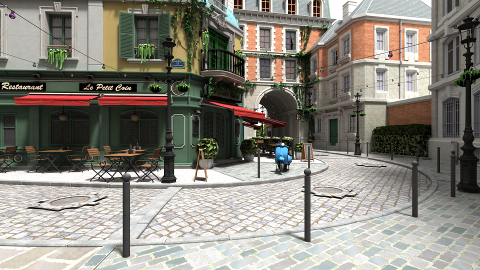}} &\\
\#triangles & 262k & 281k & 376k & 580k & 1220k & 1347k & 2833k & avg. speedup\\
\hline
\multicolumn{9}{c}{Trace Speed [MRays/s] (secondary / shadow) in DirectX}\\
\hline
Unsorted & 1755 (1.00) / 3113 (1.00) & 2228 (1.00) / 3651 (1.00) & 1904 (1.00) / 4119 (1.00) & 2355 (1.00) / 4345 (1.00) & 2436 (1.00) / 4123 (1.00) & 1086 (1.00) / 2347 (1.00) & 655 (1.00) / 1805 (1.00) & 1.00 / 1.00\\
Origin & 2826 (1.61) / 4981 (1.60) & 3539 (1.59) / 5663 (1.55) & 2794 (1.47) / 5310 (1.29) & 3500 (1.49) / 6476 (1.49) & 3305 (1.36) / 6303 (1.53) & 1478 (1.36) / 3084 (1.31) & 1346 (2.06) / \textbf{3517 (1.95)} & 1.56 / 1.53\\
Reis & 2841 (1.62) / 5000 (1.61) & 3537 (1.59) / 5703 (1.56) & \textbf{2844 (1.49)} / 5395 (1.31) & 3493 (1.48) / 6525 (1.50) & 3287 (1.35) / 6315 (1.53) & 1490 (1.37) / 3089 (1.32) & \textbf{1379 (2.11)} / 3494 (1.94) & 1.57 / \textbf{1.54}\\
Costa & 2322 (1.32) / 4812 (1.55) & 2815 (1.26) / 5552 (1.52) & 2130 (1.12) / 5029 (1.22) & 2953 (1.25) / 6356 (1.46) & 2631 (1.08) / 6123 (1.49) & 1471 (1.36) / \textbf{3132 (1.33)} & 733 (1.12) / 3286 (1.82) & 1.22 / 1.48\\
Aila & 2950 (1.68) / 5045 (1.62) & 3683 (1.65) / 5750 (1.58) & 2740 (1.44) / 5346 (1.30) & 3593 (1.53) / 6590 (1.52) & \textbf{3371 (1.38)} / 6350 (1.54) & 1611 (1.48) / 3092 (1.32) & 1289 (1.97) / 3335 (1.85) & 1.59 / 1.53\\
Aila Compact & 2977 (1.70) / \textbf{5072 (1.63)} & 3699 (1.66) / 5756 (1.58) & 2784 (1.46) / 5414 (1.31) & 3606 (1.53) / 6554 (1.51) & 3346 (1.37) / 6314 (1.53) & 1616 (1.49) / 3105 (1.32) & 1324 (2.02) / 3452 (1.91) & 1.60 / \textbf{1.54}\\
Octahedron & 2900 (1.65) / 5018 (1.61) & 3617 (1.62) / 5719 (1.57) & \textbf{2844 (1.49)} / 5392 (1.31) & 3539 (1.50) / 6503 (1.50) & 3348 (1.37) / 6286 (1.52) & 1556 (1.43) / 3105 (1.32) & 1349 (2.06) / 3498 (1.94) & 1.59 / \textbf{1.54}\\
Two Point 0.25 & 2949 (1.68) / 4940 (1.59) & \textbf{3796 (1.70)} / \textbf{5783 (1.58)} & 2656 (1.40) / 5169 (1.26) & 3615 (1.53) / 6574 (1.51) & 3322 (1.36) / 6330 (1.54) & 1637 (1.51) / 3088 (1.32) & 1134 (1.73) / 3276 (1.81) & 1.56 / 1.51\\
Two Point Adapt. & \textbf{3033 (1.73)} / 5020 (1.61) & 3793 (1.70) / 5768 (1.58) & 2801 (1.47) / \textbf{5417 (1.32)} & \textbf{3709 (1.57)} / \textbf{6592 (1.52)} & 3330 (1.37) / \textbf{6419 (1.56)} & \textbf{1654 (1.52)} / 3099 (1.32) & 1329 (2.03) / 3404 (1.89) & \textbf{1.63} / \textbf{1.54}\\
\hline
Two Point Real & 3145 (1.79) / 5000 (1.61) & 4042 (1.81) / 5740 (1.57) & 2916 (1.53) / 5401 (1.31) & 3914 (1.66) / 6567 (1.51) & 3471 (1.42) / 6371 (1.55) & 1676 (1.54) / 3077 (1.31) & 1452 (2.22) / 3416 (1.89) & 1.71 / 1.54\\
\hline
\multicolumn{9}{c}{Trace Speed [MRays/s] (secondary / shadow) in OptiX}\\
\hline
Unsorted & 1575 (1.00) / 3049 (1.00) & 1958 (1.00) / 3666 (1.00) & 1503 (1.00) / 3551 (1.00) & 2116 (1.00) / 4319 (1.00) & 2082 (1.00) / 3782 (1.00) & 867 (1.00) / 2197 (1.00) & 591 (1.00) / 1774 (1.00) & 1.00 / 1.00\\
Origin & 2384 (1.51) / 5160 (1.69) & 2936 (1.50) / 5862 (1.60) & 2048 (1.36) / 4888 (1.38) & 2950 (1.39) / 6582 (1.52) & 2646 (1.27) / 6122 (1.62) & 1120 (1.29) / 2921 (1.33) & 1075 (1.82) / \textbf{3376 (1.90)} & 1.45 / \textbf{1.58}\\
Reis & 2375 (1.51) / \textbf{5236 (1.72)} & 2868 (1.46) / 5852 (1.60) & \textbf{2156 (1.43)} / 4932 (1.39) & 2918 (1.38) / 6692 (1.55) & 2650 (1.27) / 6072 (1.61) & 1130 (1.30) / 2913 (1.33) & \textbf{1102 (1.86)} / 3321 (1.87) & 1.46 / \textbf{1.58}\\
Costa & 2045 (1.30) / 5014 (1.64) & 2430 (1.24) / 5713 (1.56) & 1677 (1.12) / 4607 (1.30) & 2660 (1.26) / 6518 (1.51) & 2276 (1.09) / 5891 (1.56) & 1183 (1.37) / \textbf{2969 (1.35)} & 642 (1.09) / 3173 (1.79) & 1.21 / 1.53\\
Aila & 2478 (1.57) / 5190 (1.70) & 3042 (1.55) / 5921 (1.62) & 2029 (1.35) / 4850 (1.37) & 3004 (1.42) / 6552 (1.52) & \textbf{2890 (1.39)} / \textbf{6186 (1.64)} & 1208 (1.39) / 2908 (1.32) & 1052 (1.78) / 3159 (1.78) & 1.49 / 1.56\\
Aila Compact & 2470 (1.57) / 5036 (1.65) & 3111 (1.59) / \textbf{6020 (1.64)} & 2066 (1.37) / \textbf{4957 (1.40)} & 3028 (1.43) / 6599 (1.53) & 2761 (1.33) / 6128 (1.62) & 1194 (1.38) / 2917 (1.33) & 1101 (1.86) / 3336 (1.88) & 1.50 / \textbf{1.58}\\
Octahedron & 2422 (1.54) / 5200 (1.71) & 2926 (1.49) / 5883 (1.60) & 2106 (1.40) / 4923 (1.39) & 2999 (1.42) / 6611 (1.53) & 2738 (1.31) / 5982 (1.58) & 1174 (1.35) / 2944 (1.34) & 1094 (1.85) / 3328 (1.88) & 1.48 / \textbf{1.58}\\
Two Point 0.25 & 2490 (1.58) / 5096 (1.67) & 3068 (1.57) / 5942 (1.62) & 1972 (1.31) / 4753 (1.34) & 3037 (1.44) / \textbf{6716 (1.55)} & 2742 (1.32) / 6016 (1.59) & 1227 (1.42) / 2903 (1.32) & 954 (1.61) / 3136 (1.77) & 1.46 / 1.55\\
Two Point Adapt. & \textbf{2590 (1.64)} / 5181 (1.70) & \textbf{3152 (1.61)} / 5937 (1.62) & 2036 (1.36) / 4934 (1.39) & \textbf{3101 (1.47)} / 6412 (1.48) & 2772 (1.33) / 6038 (1.60) & \textbf{1238 (1.43)} / 2920 (1.33) & 1092 (1.85) / 3247 (1.83) & \textbf{1.53} / 1.56\\
\hline
Two Point Real & 2649 (1.68) / 5150 (1.69) & 3286 (1.68) / 5924 (1.62) & 2109 (1.40) / 4861 (1.37) & 3247 (1.53) / 6545 (1.52) & 2880 (1.38) / 6093 (1.61) & 1259 (1.45) / 2855 (1.30) & 1183 (2.00) / 3207 (1.81) & 1.59 / 1.56\\
\hline
\hline
\end{tabular}
}
\end{center}
\label{Tab:Overview}
\end{table*}

\subsection{Ray Tracing Performance}
Table~\ref{Tab:Overview} shows a summary of the trace performance results using 32-bit sorting keys. The Two Point Real method for sorting key computation uses the trace kernel for the termination point estimation. While not practical, due to the necessity to run the trace kernel twice, it gives us the theoretical performance achievable by the Two Point method for the hypothetical case when the termination estimation would match the real result.

\begin{figure*}[htb]
  \centering
  \hfill
\includegraphics[width=\plotwidth\linewidth, trim={0.3cm 1.9cm 0.5cm 1.4cm}, clip]{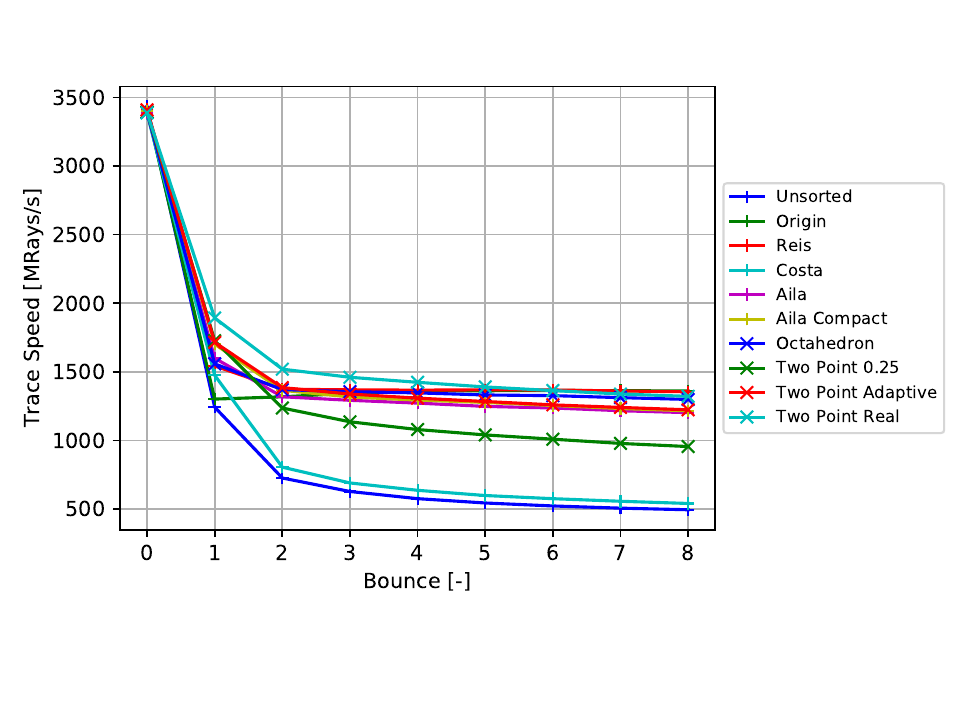}
  \hfill
\includegraphics[width=\plotwidth\linewidth, trim={0.3cm 1.9cm 0.5cm 1.4cm}, clip]{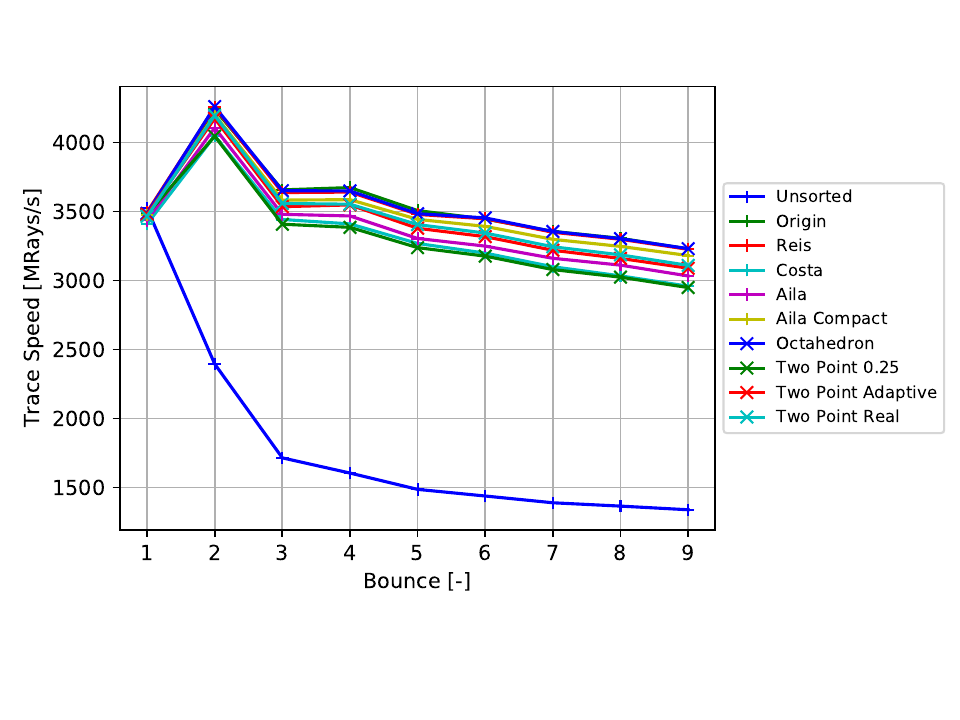}  \hfill
   \caption{\label{fig:performance-selected} The trace performance for the DirectX kernel on the Bistro scene for subsequent bounces. Primary and secondary rays (left), shadow rays (right). }
   \compress
\end{figure*}

The ray reordering leads to significant speedups of the trace kernel. For DirectX, the methods achieving the highest average speedup are: Two Point Adaptive $1.63\times$, Aila Compact $1.6\times$, Aila $1.59\times$, and Octahedron $1.59\times$ for secondary rays; Two Point Adaptive $1.54\times$, Octahedron $1.54\times$, Aila Compact $1.54\times$, and Reis $1.54\times$ for shadow rays. For OptiX, the methods achieving the highest average speedup are: Two Point Adaptive $1.53\times$, Aila Compact $1.5\times$, and Aila $1.49\times$ for secondary rays; Octahedron $1.58\times$, Reis $1.54\times$, Origin $1.54\times$, and Aila Compact $1.54\times$ for shadow rays. An example of the influence of reordering on different bounces is shown in Figure~\ref{fig:performance-selected}.

The Two Point Real method defines the current limits of ray reordering for secondary rays. It provides the average speedup of $1.71\times$ for DirectX and $1.59\times$ for OptiX. Surprisingly, for shadow rays other techniques (Octahedron, Reis, Aila) perform better than the Two Point method, probably thanks to higher directional coherence of shadow rays.

The adaptive termination point estimation strategy yields the best results overall (excluding the idealized Two Point Real method). Note that the management of the hash table of ray lengths causes only a marginal overhead. Cases when the termination point estimation fails cause traversal divergence for rays within the same thread group. The resulting decrease in trace performance is on average about 8\% for DirectX and 6\% for OptiX and is observed as a higher difference between the performance of the Two Point Real and Two Point Adaptive (see the last two rows of Table~\ref{Tab:Overview}).

\begin{figure*}[htb]
  \centering
  \hfill
\includegraphics[width=\plotwidth\linewidth, trim={0.3cm 1.9cm 0.5cm 1.4cm}, clip]{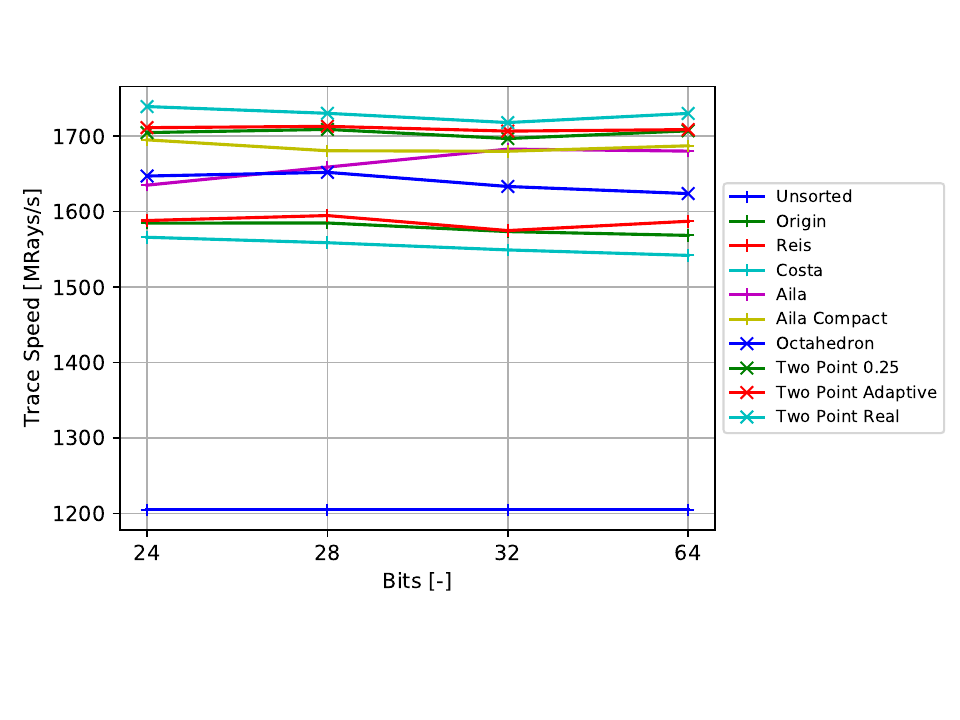}
  \hfill
\includegraphics[width=\plotwidth\linewidth, trim={0.3cm 1.9cm 0.5cm 1.4cm}, clip]{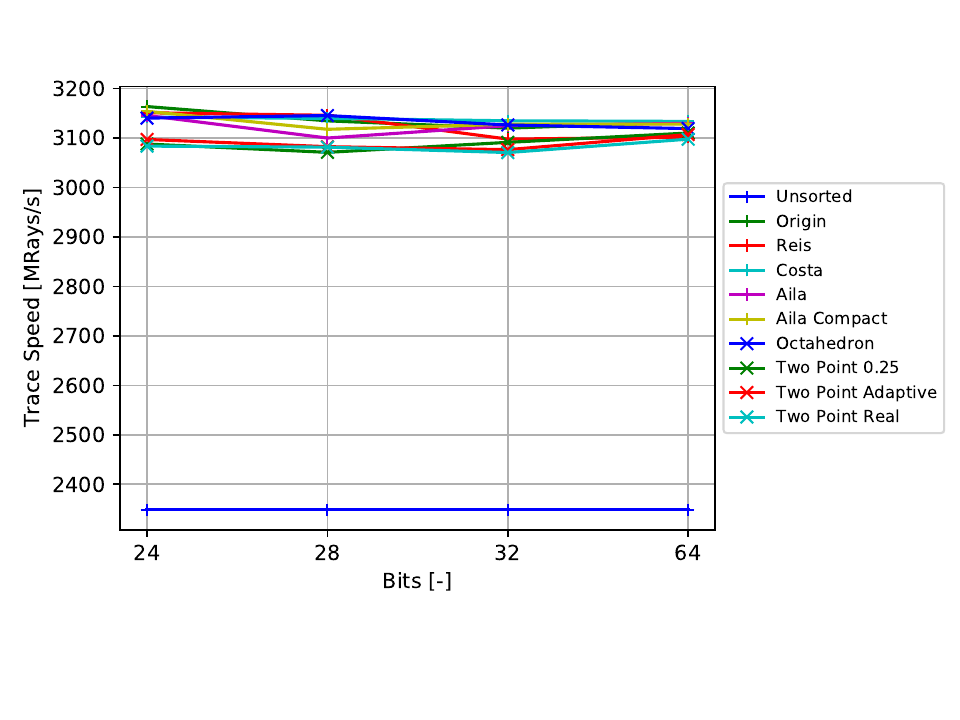}
  \hfill \\
  \hfill
\includegraphics[width=\plotwidth\linewidth, trim={0.3cm 1.9cm 0.5cm 1.4cm}, clip]{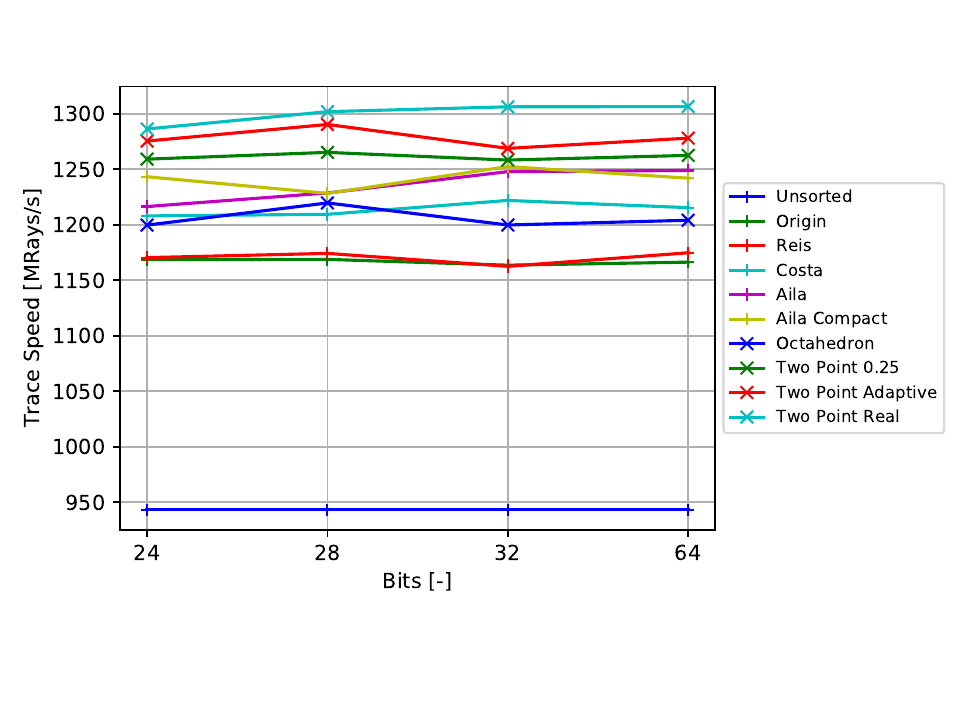}
  \hfill
\includegraphics[width=\plotwidth\linewidth, trim={0.3cm 1.9cm 0.5cm 1.4cm}, clip]{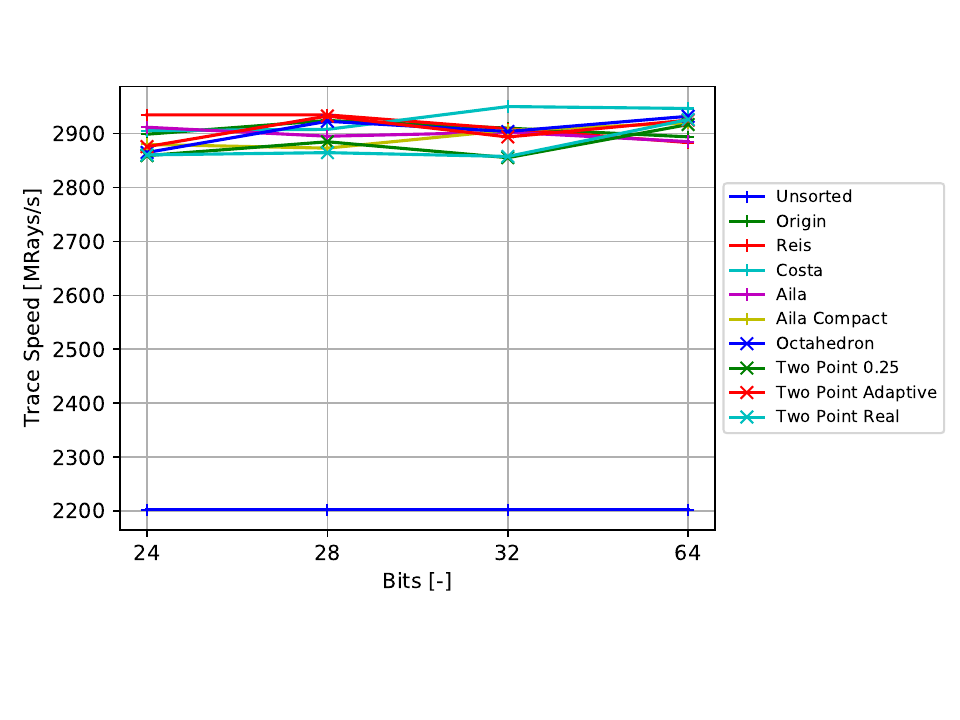}
  \hfill
   \caption{\label{fig:trace-performance} Trace performance for DirectX (top) and OptiX (bottom) kernels using sorting codes of different bit lengths on the Breakfast scene. The left column shows the trace speed for secondary rays processed using nearest-hit kernel, and the right column shows shadow rays processed using any-hit kernel.}
\compress
\end{figure*}

The behavior of the tested methods when using sorting keys of different bit-lengths is shown in Figure~\ref{fig:trace-performance}. The choice of bit-length has a significant impact on sorting performance (as discussed in Section~\ref{sec:sorting}) but does not impact the resulting ray coherence in a way to justify using longer keys. Using 64-bit keys instead of 32-bit keys results in 2.5$\times$ longer sort times, but has a marginal effect on the trace performance in all tested scenes.

\subsection{Profiling}
\begin{figure*}[htb]
  \centering
  \hfill
  \includegraphics[width=\plotwidth\linewidth]{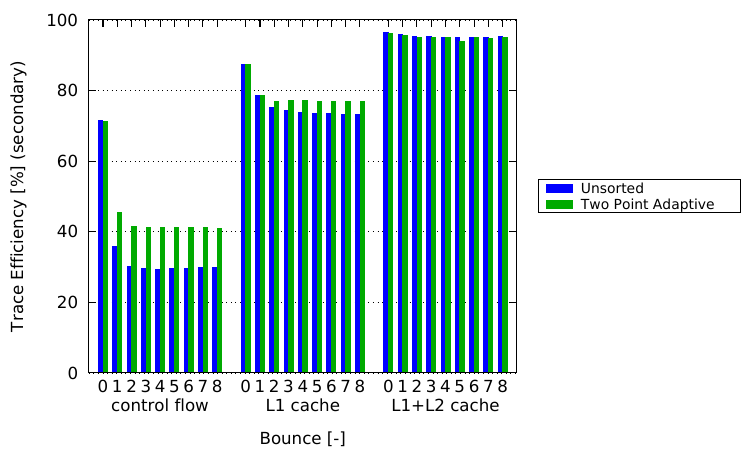}
  \hfill
  \includegraphics[width=\plotwidth\linewidth]{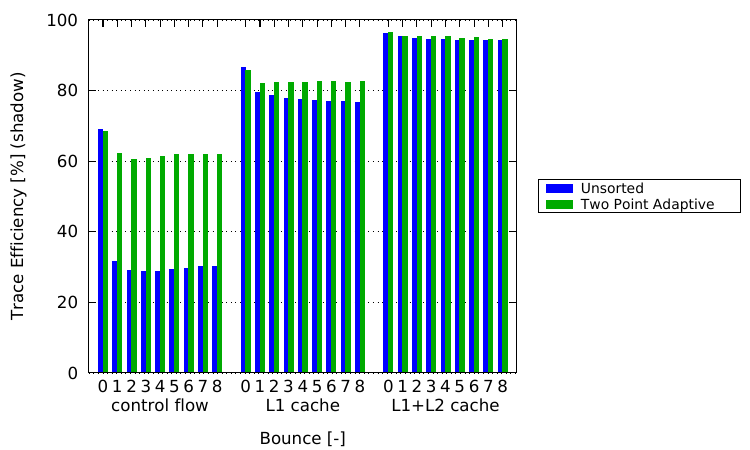}
  \hfill
   \caption{\label{fig:analysis} Control flow and cache efficiency of unsorted and sorted secondary rays (left) and shadow rays (right) for up to 8 bounces with the CUDA trace kernel in the Salle de Bain scene.}
\compress
\end{figure*}

As the main goals of increasing ray coherence are to increase control flow and cache efficiency, we performed a detailed comparison for unsorted and sorted rays using the Two Point Adaptive method by profiling the CUDA trace kernel \cite{Aila2009,Aila2012} \rev{using the NVIDIA Visual Profiler}. Figure~\ref{fig:analysis} shows the efficiency for secondary and shadow rays with up to 8 bounces for the Salle de Bain scene. The control flow efficiency increases by up to 41\% for secondary rays and up to 112\% for shadow rays, which also results in fewer memory requests for both global and local memory. In addition, the L1 cache efficiency increases by up to 5\% for secondary rays and up to 8\% for shadow rays. Because of more efficient caching in the L2 cache, the total cache efficiency for unsorted rays is almost identical to sorted rays. The total memory bandwidth is, however, up to 28\% lower for secondary rays and up to 60\% lower for shadow rays due to the higher control flow efficiency. Since the trace kernel is known to be mostly bandwidth bound, this reduction is the main source for performance improvement.

\subsection{Evaluating Ray Coherence}
Evaluating ray coherence is a complex problem. One of the possible approaches is to evaluate the coherence of decisions when traversing a particular spatial data structure~\cite{Mansson2007}. This idea is explicitly used by the streaming ray intersection techniques, which perform ray reordering on the fly~\cite{Wald2007,Barringer2014}. However, if on the fly ray reordering is complicated, or if the spatial data structure is not accessible, we rely on other means of measuring ray coherence.

One possibility is to use the surface area of the convex hull of given ray subset as a measure of its coherence. Analogously to the surface area heuristic~\cite{Goldsmith1987}, the surface area of the convex hull of the ray set is approximately proportional to the probability of the ray subset intersecting different bounding volumes. Thus, the surface area of the convex hull will correspond to the traversal footprint of a given ray set in the spatial data structure.

Due to the computational overhead and stability issues of evaluating the 3D convex hull, we propose to use a looser and simpler bounding volume for a given ray set inspired by the previous work~\cite{Szecsi2006,Roger2007}. Our ray coherence measure uses a surface area of a fitted capsule. The capsule consists of a union of two hemispheres and a conical section between them (see Figures~\ref{fig:teaser} and~\ref{fig:visualization}). Its parameters are computed as follows: we first determine the axis of the cone by computing the centroid of origins and termination points. Then we compute the average distance of the origins and termination points from this axis and use these values as radii of the corresponding hemispheres as well as the conical section between them.

The coherence of the whole ray set is evaluated by computing the average measure of ray subsets of $n$ consecutive rays. We used $n=64$ in our measurements, as our tests indicate that the RTX trace kernels processes groups of $64$ rays in parallel. We denote the corresponding measure by $M^{64}_{CPS}$ and its mean value for the whole ray set by \MCPS.

We evaluated the correlation of the proposed coherence measures with trace times for various scenes, reordering methods, and path tracing bounces. To reduce the influence of scene dependency, we used relative coherent measures and relative trace times, i.e. both quantities are normalized by the measures and trace times for the unsorted rays for each measured trace kernel call.

\begin{figure*}[htb]
  \centering
  \hfill
\includegraphics[width=\plotwidth\linewidth, trim={0.3cm 1.9cm 0.5cm 1.4cm}, clip]{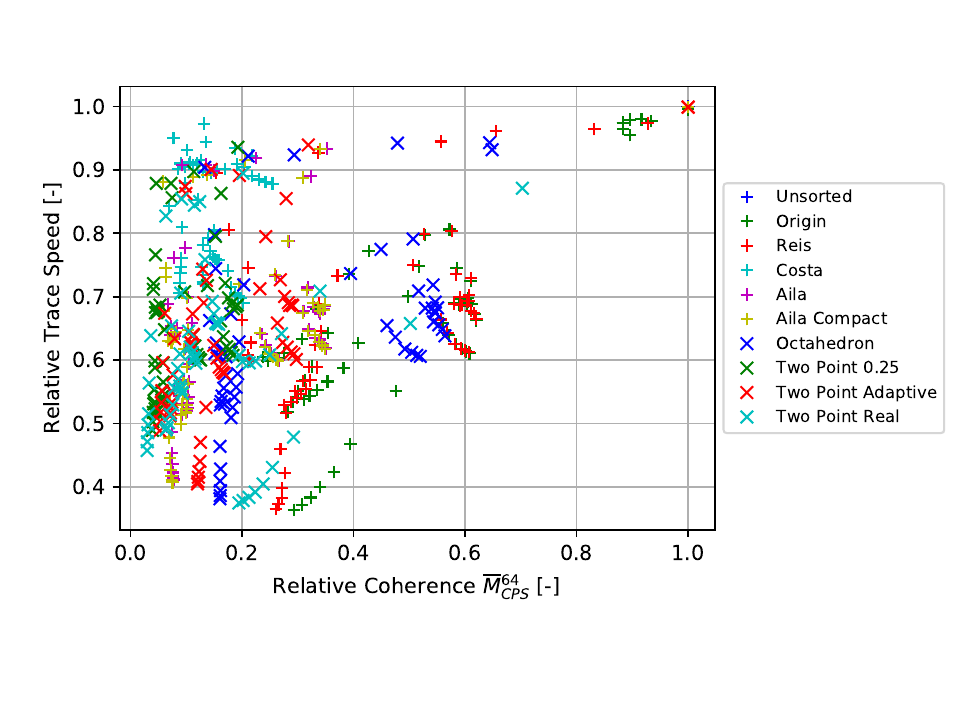}
  \hfill
\includegraphics[width=\plotwidth\linewidth, trim={0.3cm 1.9cm 0.5cm 1.4cm}, clip]{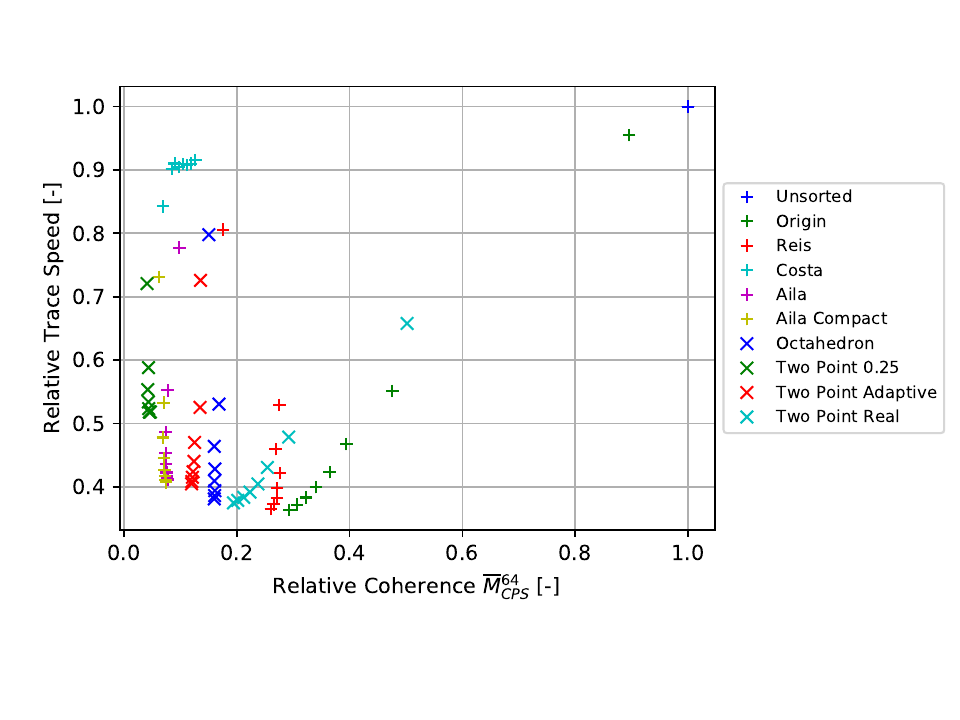}
  \hfill
   \caption{\label{fig:measures} Scatter plot of the correlation between relative coherence measure \MCPS and relative trace times. Both values are taken w.r.t. the Unsorted method: All scenes (left) and only the Bistro scene (right).}
\end{figure*}

The overall correlation for all measured scenes and methods is shown in Figure~\ref{fig:measures} on the left. The correlation for one selected scene (Bistro) is shown in Figure~\ref{fig:measures} on the right. We can observe weak to moderate overall correlation of \MCPS with trace times and a slightly stronger correlation for the single scene case. The Pearson correlation coefficient was between $0.59$ and $0.81$ for secondary rays and $0.82$ and $0.95$ for shadow rays. For the secondary rays, the lowest correlations appeared in the Salle de Bain scene, for the shadow rays in the Resort scene. The overall correlation coefficient (for all scenes and methods put together) is $0.73$ for secondary rays and $0.82$ for shadow rays.

\subsection{Discussion and Limitations}

\subsubsection{Reordering Overhead}
Although significant trace speedups were achieved, the measurements indicate, that in general ray reordering does not pay off overall due to the disproportion between the very fast trace kernel and comparatively slow ray index sorting and ray data reordering. The overhead of sorting is shown in Table~\ref{Tab:SortTimes}. We can see that the most time-consuming step is the actual ray data reordering, which uses \rev{incoherent memory access} to create coherent ray buffers. Reducing this overhead would be a key to provide a practical benefit of reordering for the trace phase for the medium size scenes that we tested. \rev{With increasing scene complexity, the reordering overhead becomes marginal in comparison with trace time as trace time increases, yet the reordering overhead remains constant.} We assume that for rendering very large scenes with complex detailed geometry, the relative overhead of reordering should be possible to recover even when combined with the fast RTX platform.

\rev{In our previous CUDA implementation, we avoided reordering by modifying the trace kernel to use indirect ray access with sorted ray index buffer. We achieved overall speedup for all tested scenes thanks to the low overhead of the indirect access and relatively slower trace kernel. For DirectX and OptiX implementation, we tried a similar approach using ray generation shaders, but then the trace speed was practically the same as for the unsorted rays. Further analysis of this behavior is an interesting topic for future work.}

\begin{table}[htb]
\caption{\rev{Breakdown of times for different phases of ray reordering, trace time and speed (secondary / shadow) for the Bistro scene. We use two different sorting strategies: global sorting (device-wise) and local sorting (block-wise) utilizing specialized kernels of the CUB library.} The times were measured using the Two Point Adaptive method in OptiX.}
\begin{center}
\scalebox{0.9}{
\begin{tabular}{l|ccc|c}
block size & 1024 & 2048 & 4096 & device-wise\\
\hline
code time [ms] & 0.34 & 0.34 & 0.34 & 0.33\\
sort time [ms] & 0.18 & 0.21 & 0.27 & 0.62\\
reorder time [ms] & 0.47 & 0.54 & 0.55 & 2.26\\
accum. time [ms] & 0.43 & 0.42 & 0.42 & 0.45\\
\hline
total overhead [ms] & 1.43 & 1.50 & 1.59 & 3.66\\
\hline
\rev{trace time [ms]} & \rev{2.24 / 1.55} & \rev{2.16 / 1.57} & \rev{2.04 / 1.51} & \rev{1.17 / 0.87}\\
\hline
trace speedup & 1.03 / 1.05 & 1.04 / 1.05 & 1.12 / 1.11 & 1.85 / 1.82\\
\end{tabular}
}
\end{center}
\label{Tab:SortTimes}
\end{table}

\subsubsection{Ray Coherence Measure}
There are several issues that prevent achieving higher correlation of the proposed ray coherence measure with the trace times. First, since the measure does not explicitly consider ray direction, it might underestimate traversal divergence, e.g.~in the case of rays with opposite directions. Second, for less dense ray distributions, the rays sample a volume defined by the convex hull sparsely, and therefore the convex hull (or any other bounding volume) is a too conservative estimate of the traversed volume. Third, defining the measure as a mean over fixed-size ray subsets does not suffice, as many ray subsets are being processed by the trace kernel. The total trace time will include non-trivial dependencies defined by the thread scheduling process and cache utility, which are not explicitly modeled by the current measure. Thus, finding a better coherence measure is an interesting research topic that might also reveal novel ray reordering algorithms.

\rev{\subsubsection{Influence on Shading Performance}
When performing ray reordering, the shader execution is more coherent, but the memory writes to the framebuffer are incoherent. Overall, the ray reordering causes higher shading times since all our test scenes have very simple materials. Similarly to the reordering overhead discussion above, we believe that for rendering scenes with more complex materials, the framebuffer scattered write overhead becomes insignificant in comparison with the benefit of coherent material accesses.}

\section{Conclusion and Future Work}
We summarized existing ray reordering techniques suitable for GPU ray tracing that are agnostic to the spatial data structure or trace kernel. We formulated a coherence measure that evaluates ray coherence using the surface area of the corresponding spatial volume. Motivated by minimizing ray incoherence, we proposed a novel method for reordering rays using estimated termination points.

We evaluated nine different ray reordering methods within DirectX and OptiX driven by the RTX technology on seven test scenes. We achieved $1.4-2.0\times$ trace speedup for DirectX and $1.3-1.9\times$ trace speedup for OptiX. The proposed Two Point ray reordering method performed the best for highly incoherent secondary rays. For shadow rays, other techniques such as the method proposed by \citet{Aila2010} or the Octahedron method~\cite{Meyer2010} work better. Overall, the results indicate a great potential of ray reordering as a preprocessing step prior to the trace phase.  Due to comparatively high overhead of ray reordering phase w.r.t. RTX accelerated trace kernels, we were not able to recover the sorting overhead even while using the state-of-the-art implementation of the sorting algorithm. 

There is a number of possible directions for future work. The current discrepancy of the results between secondary and shadow rays indicates the potential for defining other coherence measures, e.g. to put more weight on the proximity of ray origins. This could be then directly reflected in the design of improved ray reordering techniques. Another possibility is to decide on reordering strategy using a quick analysis of the input ray set. It is also possible to use other algorithms such as fast hierarchical clustering to reorder the input rays. An orthogonal research direction is to accelerate ray reordering by designing hardware sorting units, which might also be beneficial for the construction of underlying spatial acceleration structures based on sorting along a space-filling curve.

\begin{acks}
\rev{We would like to thank Benedikt Bitterli and Morgan McGuire for providing the test scenes. This research was supported by the Czech Science Foundation under project GA18-20374S and by the Research Center for Informatics No. CZ.02.1.01/0.0/0.0/16\_019/0000765.}
\end{acks}

\bibliographystyle{ACM-Reference-Format}
\bibliography{reference}


\begin{thebibliography}{34}


\ifx \showCODEN    \undefined \def \showCODEN     #1{\unskip}     \fi
\ifx \showDOI      \undefined \def \showDOI       #1{#1}\fi
\ifx \showISBNx    \undefined \def \showISBNx     #1{\unskip}     \fi
\ifx \showISBNxiii \undefined \def \showISBNxiii  #1{\unskip}     \fi
\ifx \showISSN     \undefined \def \showISSN      #1{\unskip}     \fi
\ifx \showLCCN     \undefined \def \showLCCN      #1{\unskip}     \fi
\ifx \shownote     \undefined \def \shownote      #1{#1}          \fi
\ifx \showarticletitle \undefined \def \showarticletitle #1{#1}   \fi
\ifx \showURL      \undefined \def \showURL       {\relax}        \fi
\providecommand\bibfield[2]{#2}
\providecommand\bibinfo[2]{#2}
\providecommand\natexlab[1]{#1}
\providecommand\showeprint[2][]{arXiv:#2}

\bibitem[\protect\citeauthoryear{Aila and Karras}{Aila and Karras}{2010}]%
        {Aila2010}
\bibfield{author}{\bibinfo{person}{Timo Aila} {and} \bibinfo{person}{Tero
  Karras}.} \bibinfo{year}{2010}\natexlab{}.
\newblock \showarticletitle{{Architecture Considerations for Tracing Incoherent
  Rays}}. In \bibinfo{booktitle}{\emph{Proceedings of High Performance
  Graphics}}. \bibinfo{pages}{113--122}.
\newblock


\bibitem[\protect\citeauthoryear{Aila and Laine}{Aila and Laine}{2009}]%
        {Aila2009}
\bibfield{author}{\bibinfo{person}{Timo Aila} {and} \bibinfo{person}{Samuli
  Laine}.} \bibinfo{year}{2009}\natexlab{}.
\newblock \showarticletitle{{Understanding the Efficiency of Ray Traversal on
  GPUs}}. In \bibinfo{booktitle}{\emph{Proceedings of High Performance
  Graphics}}. \bibinfo{pages}{145--149}.
\newblock


\bibitem[\protect\citeauthoryear{Aila, Laine, and Karras}{Aila
  et~al\mbox{.}}{2012}]%
        {Aila2012}
\bibfield{author}{\bibinfo{person}{Timo Aila}, \bibinfo{person}{Samuli Laine},
  {and} \bibinfo{person}{Tero Karras}.} \bibinfo{year}{2012}\natexlab{}.
\newblock \showarticletitle{{Understanding the Efficiency of Ray Traversal on
  GPUs -- Kepler and Fermi Addendum}}. In \bibinfo{booktitle}{\emph{Proceedings
  of High Performance Graphics (Posters)}}. \bibinfo{pages}{9--16}.
\newblock


\bibitem[\protect\citeauthoryear{Arvo and Kirk}{Arvo and Kirk}{1987}]%
        {Arvo1987}
\bibfield{author}{\bibinfo{person}{James Arvo} {and} \bibinfo{person}{David
  Kirk}.} \bibinfo{year}{1987}\natexlab{}.
\newblock \showarticletitle{{Fast Ray Tracing by Ray Classification}}. In
  \bibinfo{booktitle}{\emph{Proceedings of International Conference on Computer
  Graphics and Interactive Techniques}}. \bibinfo{pages}{55--64}.
\newblock


\bibitem[\protect\citeauthoryear{Barringer and Akenine-M\"{o}ller}{Barringer
  and Akenine-M\"{o}ller}{2014}]%
        {Barringer2014}
\bibfield{author}{\bibinfo{person}{Rasmus Barringer} {and}
  \bibinfo{person}{Tomas Akenine-M\"{o}ller}.} \bibinfo{year}{2014}\natexlab{}.
\newblock \showarticletitle{{Dynamic Ray Stream Traversal}}.
\newblock \bibinfo{journal}{\emph{ACM Transactions on Graphics}}
  \bibinfo{volume}{33}, \bibinfo{number}{4} (\bibinfo{year}{2014}),
  \bibinfo{pages}{151:1--151:9}.
\newblock


\bibitem[\protect\citeauthoryear{Bikker}{Bikker}{2012}]%
        {Bikker2012}
\bibfield{author}{\bibinfo{person}{Jacco Bikker}.}
  \bibinfo{year}{2012}\natexlab{}.
\newblock \showarticletitle{{Improving Data Locality for Efficient In-Core Path
  Tracing}}.
\newblock \bibinfo{journal}{\emph{Computer Graphics Forum}}
  \bibinfo{volume}{31}, \bibinfo{number}{6} (\bibinfo{year}{2012}),
  \bibinfo{pages}{1936--1947}.
\newblock


\bibitem[\protect\citeauthoryear{Bitterli}{Bitterli}{2016}]%
        {Bitterli2016}
\bibfield{author}{\bibinfo{person}{Benedikt Bitterli}.}
  \bibinfo{year}{2016}\natexlab{}.
\newblock \bibinfo{title}{{Rendering Resources}}.
\newblock
\newblock
\newblock
\shownote{https://benedikt-bitterli.me/resources/.}


\bibitem[\protect\citeauthoryear{Boulos, Wald, and Benthin}{Boulos
  et~al\mbox{.}}{2008}]%
        {Boulos2008}
\bibfield{author}{\bibinfo{person}{Solomon Boulos}, \bibinfo{person}{Ingo
  Wald}, {and} \bibinfo{person}{Carsten Benthin}.}
  \bibinfo{year}{2008}\natexlab{}.
\newblock \showarticletitle{{Adaptive Ray Packet Reordering}}. In
  \bibinfo{booktitle}{\emph{Proceedings of IEEE Symposium on Interactive Ray
  Tracing}}.
\newblock


\bibitem[\protect\citeauthoryear{Costa, Pereira, and Jorge}{Costa
  et~al\mbox{.}}{2015}]%
        {Costa2015}
\bibfield{author}{\bibinfo{person}{Vasco Costa},
  \bibinfo{person}{Jo\~{a}o~Madeiras Pereira}, {and}
  \bibinfo{person}{Joaquim~A. Jorge}.} \bibinfo{year}{2015}\natexlab{}.
\newblock \showarticletitle{{Accelerating Occlusion Rendering on a GPU via Ray
  Classification}}.
\newblock \bibinfo{journal}{\emph{Int. Journal of Creative Interfaces and Comp.
  Graphics}} \bibinfo{volume}{6}, \bibinfo{number}{2} (\bibinfo{year}{2015}),
  \bibinfo{pages}{1--17}.
\newblock
\showISSN{1947-3117}


\bibitem[\protect\citeauthoryear{Eisenacher, Nichols, Selle, and
  Burley}{Eisenacher et~al\mbox{.}}{2013}]%
        {Eisenacher2013}
\bibfield{author}{\bibinfo{person}{Christian Eisenacher},
  \bibinfo{person}{Gregory Nichols}, \bibinfo{person}{Andrew Selle}, {and}
  \bibinfo{person}{Brent Burley}.} \bibinfo{year}{2013}\natexlab{}.
\newblock \showarticletitle{{Sorted Deferred Shading for Production Path
  Tracing}}.
\newblock \bibinfo{journal}{\emph{Computer Graphics Forum}}
  \bibinfo{volume}{32}, \bibinfo{number}{4} (\bibinfo{year}{2013}),
  \bibinfo{pages}{125--132}.
\newblock


\bibitem[\protect\citeauthoryear{Garanzha and Loop}{Garanzha and Loop}{2010}]%
        {Garanzha2010}
\bibfield{author}{\bibinfo{person}{Kirill Garanzha} {and}
  \bibinfo{person}{Charles Loop}.} \bibinfo{year}{2010}\natexlab{}.
\newblock \showarticletitle{{Fast Ray Sorting and Breadth-First Packet
  Traversal for GPU Ray Tracing}}.
\newblock \bibinfo{journal}{\emph{Computer Graphics Forum}}
  \bibinfo{volume}{29}, \bibinfo{number}{2} (\bibinfo{year}{2010}),
  \bibinfo{pages}{289--298}.
\newblock


\bibitem[\protect\citeauthoryear{Goldsmith and Salmon}{Goldsmith and
  Salmon}{1987}]%
        {Goldsmith1987}
\bibfield{author}{\bibinfo{person}{Jeffrey Goldsmith} {and}
  \bibinfo{person}{John Salmon}.} \bibinfo{year}{1987}\natexlab{}.
\newblock \showarticletitle{{Automatic Creation of Object Hierarchies for Ray
  Tracing}}.
\newblock \bibinfo{journal}{\emph{Comput Graphics and Applications}}
  \bibinfo{volume}{7}, \bibinfo{number}{5} (\bibinfo{year}{1987}),
  \bibinfo{pages}{14--20}.
\newblock


\bibitem[\protect\citeauthoryear{Gunther, Popov, Seidel, and Slusallek}{Gunther
  et~al\mbox{.}}{2007}]%
        {Gunther2007}
\bibfield{author}{\bibinfo{person}{Johannes Gunther}, \bibinfo{person}{Stefan
  Popov}, \bibinfo{person}{Hans-Peter Seidel}, {and} \bibinfo{person}{Philipp
  Slusallek}.} \bibinfo{year}{2007}\natexlab{}.
\newblock \showarticletitle{{Realtime Ray Tracing on GPU with BVH-based Packet
  Traversal}}. In \bibinfo{booktitle}{\emph{Proceedings of IEEE Symposium on
  Interactive Ray Tracing}}. \bibinfo{pages}{113--118}.
\newblock


\bibitem[\protect\citeauthoryear{Hanika, Keller, and Lensch}{Hanika
  et~al\mbox{.}}{2010}]%
        {Hanika2010}
\bibfield{author}{\bibinfo{person}{Johannes Hanika}, \bibinfo{person}{Alexander
  Keller}, {and} \bibinfo{person}{Hendrik P.~A. Lensch}.}
  \bibinfo{year}{2010}\natexlab{}.
\newblock \showarticletitle{{Two-Level Ray Tracing with Reordering for Highly
  Complex Scenes}}. In \bibinfo{booktitle}{\emph{Proceedings of Graphics
  Interface 2010}}. \bibinfo{pages}{145--–152}.
\newblock


\bibitem[\protect\citeauthoryear{Laine, Karras, and Aila}{Laine
  et~al\mbox{.}}{2013}]%
        {Laine2013}
\bibfield{author}{\bibinfo{person}{Samuli Laine}, \bibinfo{person}{Tero
  Karras}, {and} \bibinfo{person}{Timo Aila}.} \bibinfo{year}{2013}\natexlab{}.
\newblock \showarticletitle{{Megakernels Considered Harmful: Wavefront Path
  Tracing on GPUs}}. In \bibinfo{booktitle}{\emph{Proceedings of High
  Performance Graphics}}. \bibinfo{pages}{137--143}.
\newblock
\showISBNx{978-1-4503-2135-8}


\bibitem[\protect\citeauthoryear{Lauterbach, Garland, Sengupta, Luebke, and
  Manocha}{Lauterbach et~al\mbox{.}}{2009}]%
        {Lauterbach2009}
\bibfield{author}{\bibinfo{person}{Christian Lauterbach},
  \bibinfo{person}{Michael Garland}, \bibinfo{person}{Shubhabrata Sengupta},
  \bibinfo{person}{David Luebke}, {and} \bibinfo{person}{Dinesh Manocha}.}
  \bibinfo{year}{2009}\natexlab{}.
\newblock \showarticletitle{Fast {BVH} Construction on {GPU}s}.
\newblock \bibinfo{journal}{\emph{Computer Graphics Forum}}
  \bibinfo{volume}{28}, \bibinfo{number}{2} (\bibinfo{year}{2009}),
  \bibinfo{pages}{375--384}.
\newblock


\bibitem[\protect\citeauthoryear{Lloyd}{Lloyd}{1982}]%
        {Lloyd1982}
\bibfield{author}{\bibinfo{person}{Stuart~P. Lloyd}.}
  \bibinfo{year}{1982}\natexlab{}.
\newblock \showarticletitle{{Least Squares Quantization in PCM}}.
\newblock \bibinfo{journal}{\emph{IEEE Transactions on Information Theory}}
  \bibinfo{volume}{28}, \bibinfo{number}{2} (\bibinfo{year}{1982}),
  \bibinfo{pages}{129--137}.
\newblock


\bibitem[\protect\citeauthoryear{Mansson, Munkberg, and Akenine-Moller}{Mansson
  et~al\mbox{.}}{2007}]%
        {Mansson2007}
\bibfield{author}{\bibinfo{person}{Erik Mansson}, \bibinfo{person}{Jacob
  Munkberg}, {and} \bibinfo{person}{Tomas Akenine-Moller}.}
  \bibinfo{year}{2007}\natexlab{}.
\newblock \showarticletitle{Deep Coherent Ray Tracing}. In
  \bibinfo{booktitle}{\emph{Proceedings of IEEE Symposium on Interactive Ray
  Tracing}}. \bibinfo{pages}{79--85}.
\newblock


\bibitem[\protect\citeauthoryear{McGuire}{McGuire}{2017}]%
        {McGuire2017}
\bibfield{author}{\bibinfo{person}{Morgan McGuire}.}
  \bibinfo{year}{2017}\natexlab{}.
\newblock \bibinfo{title}{{Computer Graphics Archive}}.
\newblock
\newblock
\newblock
\shownote{https://casual-effects.com/data.}


\bibitem[\protect\citeauthoryear{Merrill and Grimshaw}{Merrill and
  Grimshaw}{2011}]%
        {Merrill2011}
\bibfield{author}{\bibinfo{person}{Duane Merrill} {and} \bibinfo{person}{Andrew
  Grimshaw}.} \bibinfo{year}{2011}\natexlab{}.
\newblock \showarticletitle{{High Performance and Scalable Radix Sorting: A
  Case Study of Implementing Dynamic Parallelism for {GPU} Computing}}.
\newblock \bibinfo{journal}{\emph{Parallel Processing Letters}}
  \bibinfo{volume}{21}, \bibinfo{number}{02} (\bibinfo{year}{2011}),
  \bibinfo{pages}{245--272}.
\newblock


\bibitem[\protect\citeauthoryear{Meyer, Süßmuth, Sußner, Stamminger, and
  Greiner}{Meyer et~al\mbox{.}}{2010}]%
        {Meyer2010}
\bibfield{author}{\bibinfo{person}{Quirin Meyer}, \bibinfo{person}{Jochen
  Süßmuth}, \bibinfo{person}{Gerd Sußner}, \bibinfo{person}{Marc
  Stamminger}, {and} \bibinfo{person}{Günther Greiner}.}
  \bibinfo{year}{2010}\natexlab{}.
\newblock \showarticletitle{{On Floating-Point Normal Vectors}}.
\newblock \bibinfo{journal}{\emph{Computer Graphics Forum}}
  \bibinfo{volume}{29}, \bibinfo{number}{4} (\bibinfo{year}{2010}),
  \bibinfo{pages}{1405--1409}.
\newblock


\bibitem[\protect\citeauthoryear{Moon, Byun, Kim, Claudio, Kim, Ban, Nam, and
  Yoon}{Moon et~al\mbox{.}}{2010}]%
        {Moon2010}
\bibfield{author}{\bibinfo{person}{Bochang Moon}, \bibinfo{person}{Yongyoung
  Byun}, \bibinfo{person}{Tae-Joon Kim}, \bibinfo{person}{Pio Claudio},
  \bibinfo{person}{Hye-Sun Kim}, \bibinfo{person}{Yun-Ji Ban},
  \bibinfo{person}{Seung~Woo Nam}, {and} \bibinfo{person}{Sung-Eui Yoon}.}
  \bibinfo{year}{2010}\natexlab{}.
\newblock \showarticletitle{{Cache-Oblivious Ray Reordering}}.
\newblock \bibinfo{journal}{\emph{ACM Transactions on Graphics}}
  \bibinfo{volume}{29}, \bibinfo{number}{3} (\bibinfo{year}{2010}),
  \bibinfo{pages}{1--10}.
\newblock


\bibitem[\protect\citeauthoryear{Navratil, Fussell, Lin, and Mark}{Navratil
  et~al\mbox{.}}{2007}]%
        {Navratil2007}
\bibfield{author}{\bibinfo{person}{Paul~Arthur Navratil},
  \bibinfo{person}{Donald~S. Fussell}, \bibinfo{person}{Calvin Lin}, {and}
  \bibinfo{person}{William~R. Mark}.} \bibinfo{year}{2007}\natexlab{}.
\newblock \showarticletitle{{Dynamic Ray Scheduling to Improve Ray Coherence
  and Bandwidth Utilization}}. In \bibinfo{booktitle}{\emph{Proceedings of IEEE
  Symposium on Interactive Ray Tracing}}. \bibinfo{pages}{95--104}.
\newblock


\bibitem[\protect\citeauthoryear{Nimier-David, Vicini, Zeltner, and
  Jakob}{Nimier-David et~al\mbox{.}}{2019}]%
        {Nimier2019}
\bibfield{author}{\bibinfo{person}{Merlin Nimier-David}, \bibinfo{person}{Delio
  Vicini}, \bibinfo{person}{Tizian Zeltner}, {and} \bibinfo{person}{Wenzel
  Jakob}.} \bibinfo{year}{2019}\natexlab{}.
\newblock \showarticletitle{{Mitsuba 2: A Retargetable Forward and Inverse
  Renderer}}.
\newblock \bibinfo{journal}{\emph{ACM Transactions on Graphics}}
  \bibinfo{volume}{38}, \bibinfo{number}{6} (\bibinfo{year}{2019}),
  \bibinfo{pages}{203}.
\newblock


\bibitem[\protect\citeauthoryear{Pharr, Kolb, Gershbein, and Hanrahan}{Pharr
  et~al\mbox{.}}{1997}]%
        {Pharr1997}
\bibfield{author}{\bibinfo{person}{Matt Pharr}, \bibinfo{person}{Craig Kolb},
  \bibinfo{person}{Reid Gershbein}, {and} \bibinfo{person}{Pat Hanrahan}.}
  \bibinfo{year}{1997}\natexlab{}.
\newblock \showarticletitle{{Rendering Complex Scenes with Memory-coherent Ray
  Tracing}}. In \bibinfo{booktitle}{\emph{Proceedings of International
  Conference on Computer Graphics and Interactive Techniques}}.
  \bibinfo{pages}{101--108}.
\newblock


\bibitem[\protect\citeauthoryear{Reis, Costa, and Pereira}{Reis
  et~al\mbox{.}}{2017}]%
        {Reis2017}
\bibfield{author}{\bibinfo{person}{Nuno~T. Reis}, \bibinfo{person}{Vasco~S.
  Costa}, {and} \bibinfo{person}{Jo{\~a}o~M. Pereira}.}
  \bibinfo{year}{2017}\natexlab{}.
\newblock \showarticletitle{{Coherent Ray-Space Hierarchy via Ray Hashing and
  Sorting}}. In \bibinfo{booktitle}{\emph{Proceedings of International Joint
  Conference on Computer Vision, Imaging, and Computer Graphics Theory and
  Applications}}.
\newblock


\bibitem[\protect\citeauthoryear{Reshetov, Soupikov, and Hurley}{Reshetov
  et~al\mbox{.}}{2005}]%
        {Reshetov2005}
\bibfield{author}{\bibinfo{person}{Alexander Reshetov}, \bibinfo{person}{Alexei
  Soupikov}, {and} \bibinfo{person}{Jim Hurley}.}
  \bibinfo{year}{2005}\natexlab{}.
\newblock \showarticletitle{{Multi-Level Ray Tracing Algorithm}}.
\newblock \bibinfo{journal}{\emph{ACM Transactions on Graphics (Proceedings of
  ACM SIGGRAPH 2005)}} \bibinfo{volume}{24}, \bibinfo{number}{3}
  (\bibinfo{year}{2005}), \bibinfo{pages}{1176--1185}.
\newblock


\bibitem[\protect\citeauthoryear{Roger, Assarsson, and Holzschuch}{Roger
  et~al\mbox{.}}{2007}]%
        {Roger2007}
\bibfield{author}{\bibinfo{person}{David Roger}, \bibinfo{person}{Ulf
  Assarsson}, {and} \bibinfo{person}{Nicolas Holzschuch}.}
  \bibinfo{year}{2007}\natexlab{}.
\newblock \showarticletitle{{Whitted Ray-Tracing for Dynamic Scenes using a
  Ray-Space Hierarchy on the GPU}}. In \bibinfo{booktitle}{\emph{Proceedings of
  the Eurographics Symposium on Rendering Techniques}}.
  \bibinfo{pages}{99--110}.
\newblock


\bibitem[\protect\citeauthoryear{Sz{\'{e}}csi}{Sz{\'{e}}csi}{2006}]%
        {Szecsi2006}
\bibfield{author}{\bibinfo{person}{L\'{a}szl\'{o} Sz{\'{e}}csi}.}
  \bibinfo{year}{2006}\natexlab{}.
\newblock \showarticletitle{{The Hierarchical Ray Engine}}. In
  \bibinfo{booktitle}{\emph{Proceedings of International Conference in Central
  Europe on Computer Graphics, Visualization and Computer Vision}}.
  \bibinfo{pages}{249--256}.
\newblock


\bibitem[\protect\citeauthoryear{Szirmay-Kalos and Purgathofer}{Szirmay-Kalos
  and Purgathofer}{1998}]%
        {Szirmay1998}
\bibfield{author}{\bibinfo{person}{Laszlo Szirmay-Kalos} {and}
  \bibinfo{person}{Werner Purgathofer}.} \bibinfo{year}{1998}\natexlab{}.
\newblock \showarticletitle{Global Ray-Bundle Tracing with Infinite Number of
  Rays}.
\newblock \bibinfo{journal}{\emph{Computers \& Graphics}} \bibinfo{volume}{23},
  \bibinfo{number}{2} (\bibinfo{year}{1998}), \bibinfo{pages}{193--202}.
\newblock


\bibitem[\protect\citeauthoryear{van Antwerpen}{van Antwerpen}{2011}]%
        {vanAntwerpen2011}
\bibfield{author}{\bibinfo{person}{Dietger van Antwerpen}.}
  \bibinfo{year}{2011}\natexlab{}.
\newblock \showarticletitle{{Improving SIMD Efficiency for Parallel Monte Carlo
  Light Transport on the GPU}}. In \bibinfo{booktitle}{\emph{Proceedings of
  High Performance Graphics}}. 10.
\newblock


\bibitem[\protect\citeauthoryear{Wald}{Wald}{2011}]%
        {Wald2011}
\bibfield{author}{\bibinfo{person}{Ingo Wald}.}
  \bibinfo{year}{2011}\natexlab{}.
\newblock \showarticletitle{{Active Thread Compaction for GPU Path Tracing}}.
  In \bibinfo{booktitle}{\emph{Proceedings of High Performance Graphics}}.
  \bibinfo{pages}{51--58}.
\newblock


\bibitem[\protect\citeauthoryear{Wald, Gribble, Boulos, and Kensler}{Wald
  et~al\mbox{.}}{2007}]%
        {Wald2007}
\bibfield{author}{\bibinfo{person}{Ingo Wald}, \bibinfo{person}{Christiaan~P.
  Gribble}, \bibinfo{person}{Solomon Boulos}, {and} \bibinfo{person}{Andrew
  Kensler}.} \bibinfo{year}{2007}\natexlab{}.
\newblock \showarticletitle{{SIMD Ray Stream Tracing - SIMD Ray Traversal with
  Generalized Ray Packets and On-the-fly Re-Ordering}}. In
  \bibinfo{booktitle}{\emph{Technical Report UUSCI-2007-012}}.
\newblock


\bibitem[\protect\citeauthoryear{Wald, Purcell, Schmittler, Benthin, and
  Slusallek}{Wald et~al\mbox{.}}{2003}]%
        {Wald2003}
\bibfield{author}{\bibinfo{person}{Ingo Wald}, \bibinfo{person}{Timothy~J.
  Purcell}, \bibinfo{person}{Joerg Schmittler}, \bibinfo{person}{Carsten
  Benthin}, {and} \bibinfo{person}{Philipp Slusallek}.}
  \bibinfo{year}{2003}\natexlab{}.
\newblock \bibinfo{title}{{Realtime Ray Tracing and Its Use for Interactive
  Global Illumination}}.
\newblock \bibinfo{howpublished}{Eurographics State of the Art Reports}.
\newblock


\end{thebibliography}
\end{document}